\documentclass[pra,aps,twocolumn,showpacs,lengthcheck,preprintnumbers,
superscriptaddress]{revtex4-1}
\usepackage{amsfonts}
\usepackage{amsmath}
\usepackage{amssymb}
\usepackage{graphicx}
\usepackage{fontenc}
\newcommand{\floor}[1]{\lfloor #1 \rfloor}
\usepackage{color}

\begin{document}

\title[Transverse Josephson vortices and localized states in stacked 
BECs]{Transverse Josephson vortices and localized states \\ 
in stacked Bose-Einstein condensates}
\author{J. A. Gil Granados}
\affiliation{
Departament de F\'isica Qu\`antica i Astrof\'isica, 
Facultat de F\'{\i}sica, Universitat de Barcelona, E--08028
Barcelona, Spain}
\author{A. Mu\~{n}oz Mateo}
\affiliation{
Departament de F\'isica Qu\`antica i Astrof\'isica, 
Facultat de F\'{\i}sica, Universitat de Barcelona, E--08028
Barcelona, Spain}
\affiliation{
Centre for
Theoretical Chemistry and Physics, New Zealand Institute for Advanced Study,
Massey University, Private Bag 102904 NSMC, Auckland 0745, New Zealand}
\affiliation{
Dodd-Walls Centre for Photonic and Quantum Technologies, New Zealand}
\author{M. Guilleumas}
\affiliation{
Departament de F\'isica Qu\`antica i Astrof\'isica, 
Facultat de F\'{\i}sica, Universitat de Barcelona, E--08028
Barcelona, Spain}
\affiliation{
Institut de Ci\`encies del Cosmos, Universitat de Barcelona, ICCUB, 
08028-Barcelona, Spain}
\author{X. Vi\~nas}
\affiliation{
Departament de F\'isica Qu\`antica i Astrof\'isica, 
Facultat de F\'{\i}sica, Universitat de Barcelona, E--08028
Barcelona, Spain}
\affiliation{
Institut de Ci\`encies del Cosmos, Universitat de Barcelona, ICCUB, 
08028-Barcelona, Spain}


\begin{abstract}
The stacks of Bose-Einstein condensates coupled by long Josephson 
junctions present a rich phenomenology feasible to experimental 
realization and specially suitable for technological applications as the
nonlinear-optics and superconducting analogues have already proved. 
Among this, we show that transverse Bloch waves excited in 
arrays of one-dimensional coupled condensates can carry 
tunneling superflows whose dynamical stability depends on the 
quasimomentum.
Across the stacks with periodic boundary conditions, forming closed ring-shaped 
systems, such Bloch states yield transverse Josephson vortices 
with a generic non-integer circulation in units of $h/m$. 
Additionally, the superpositions of degenerate linear Bloch 
waves can suppress the supercurrents and give rise to families of nonlinear 
standing-wave states with strong (transverse) spatial localization. Stable 
states of this type can also be found in finite size systems.
\end{abstract}

\maketitle

\section{Introduction}

Quantum tunneling is one of the most striking phenomena predicted by 
quantum mechanics. At a macroscopic scale it is named  
Josephson effect, and it is a paradigm of the
phase coherence manifestation of a macroscopic quantum system.
The theory of the Josephson effect was developed by B. D. Josephson for 
superconducting electron pairs in 1962~\cite{Josephson1962}. Since then, it has 
found multiple technological applications \cite{Barone1982,Askerzade2017}. 
More recently, with the advent of Bose-Einstein 
condensates (BECs) of ultracold atoms, the Josephson effect has been
demonstrated between two weakly linked condensates of neutral bosonic atoms
{\color{black}\cite{Smerzi2003,Albiez2005,Levy2007}}. 

A bosonic, macroscopic quantum system can be described by a complex order 
parameter whose squared modulus and phase gradient provide the
particle density and the particle current, respectively.
When two such systems are connected by a weak link, that is a Josephson 
junction, the macroscopic tunneling of particles through the junction varies as 
the sine of the relative phase between the coupled order parameters.
This supercurrent can flow through point-like (or short) Josephson junctions, 
as it is the case of the barrier in a double well potential 
~\cite{Albiez2005,Levy2007}, but also through long Josephson junctions with a 
non-negligible spatial extension. In the latter case, the 
relative phase can change along the junction and the coherent transfer of 
particles occurs locally through each point. Feasible realizations
of long Josephson junctions in ultracold atomic gases can be 
readily done by Raman-laser coupling of different hyperfine components of 
atomic BECs producing a so-called internal Josephson effect 
{\color{black}~\cite{Smerzi2003,Sols1999,Williams1999b}}, or by spatially 
separated BECs coupled along their longitudinal direction, as we consider in 
the present paper.

In general, the long Josephson junctions in BECs have received less attention 
than the short ones. Most of the previous works with systems containing long 
junctions have focused on coupled binary BECs (see e.g. \cite{Abad2013} 
and references therein), and significant attention have been paid to states 
containing localized Josephson vortices or fluxons
\cite{Kaurov2005,Kaurov2006,Brand2009,Qadir2012a,Sophie2018}. These
topological structures are characteristic of the long junctions and have been 
extensively studied in superconductors because of their capability of trapping 
magnetic flux \cite{Barone1982}. They involve localized supercurrents around a 
vortex core situated in the Josephson junction \cite{Roditchev2015}, and can be 
theoretically studied as analytical solutions to the sine-Gordon equation for 
the relative phase of the coupled systems \cite{Barone1982}. In this regard, 
one-dimensional tunnel-coupled superfluids, as quantum simulator of the 
sine-Gordon equation, 
have been recently realized in ultracold gases \cite{Schweigler2017}. 
Within the mean field framework, the dynamical properties of Josephson vortices 
have been also studied in two coupled 1D BECs 
\cite{Kaurov2005,Kaurov2006,Brand2009,Qadir2012a,Sophie2018,Son2002,Qu2017}.
Generalizations of Josephson vortex states to tunnel-coupled spinor gases 
\cite{Montgomery2013} and to multidimensional spin-orbit coupled condensates 
\cite{Gallemi2016} have been proposed.

Due to their potential for technical applications, the arrays of linear and 
nonlinear coupled waveguides are the subject of intense experimental 
and theoretical research in optics (see 
e.g. \cite{Christodoulides2003,Lederer2008} and references therein), where the 
discrete nonlinear Schr\"odinger equation serves as a usual theoretical model 
of the array. 
Likewise, the stacks of long Josephson junctions have been extensively 
studied in superconductors, where the stacks can be modeled by coupled 
sine-Gordon equations (see, for instance \cite{Kivshar1988,Mazo2014}). However, 
in BECs, as 
far as we know, up to date only 1D arrays of point-like 
Josephson junctions have been experimentally realized 
\cite{Cataliotti2001}.
There is not yet a wide theoretical exploration of bosonic-array systems with 
long spatial couplings either. Nevertheless, different aspects of the theory  
have been addressed: The superfluid-insulator transition has been studied 
in two-dimensional (2D) arrays of coupled 1D tubes against the 
absence or presence of axial periodic potentials \cite{Cazalilla2006}; also, 
the propagation of bright solitons in arrays of BECs with negative nonlinearity 
has been considered \cite{Blit2012};
very recently, coupled atomic wires have been 
proposed in ultracold-gas systems for the generation of exotic phases in the 
presence of synthetic gauge 
fields \cite{Budich2017}; in addition, the arrays of parallel one-dimensional 
long Josephson junctions in BECs with positive nonlinearity have been 
demonstrated to provide an excellent playground for the realization and 
stabilization of solitary waves \cite{Baals2018}. 

In this work we study, analytically and numerically, a stack of linearly 
coupled 1D BECs with repulsive interparticle interactions that gives rise to an 
underlying array of coupled-parallel long Josephson junctions.  
In particular, we consider stacks with periodic boundary conditions forming
closed, ring-shaped arrays. By solving the 
Gross-Pitaevskii equation, the dynamics of stationary 
states composing a transverse, discrete Bloch band is addressed.
We show that the periodic boundary conditions yield
transverse vortex states that carry Josephson supercurrents. Interestingly, the 
Bogoliubov analysis reveals that, with a quasimomentum dependence, these 
Josephson vortices can be dynamically stable, hence susceptible of experimental 
detection. From a hydrodynamical 
perspective, we also perform a long-wavelength linear analysis that allow for 
a somehow simpler interpretation of the system 
dynamics in terms of coupled wave equations for the relative 
phases and densities. We demonstrate that, only for 
particular states in the small coupling limit, {\color{black} also known as the 
anti-continuum limit in the literature of discrete systems,} the resulting 
hydrodynamic 
model resembles, although it cannot be fully identified with, the 
superconducting models of coupled 
sine-Gordon equations. Finally, we explore how the linear combinations of Bloch 
waves lead to families of nonlinear states that cancel the Josephson 
supercurrents and produce strongly localized density profiles across the stack. 
In the limit of small coupling, these states can also be dynamically stable 
against perturbations in finite size systems.

The paper is organized as follows: in Sec. II we present the theoretical model 
that describes the stack of linearly coupled BECs with periodic boundary 
conditions in the mean-field framework. We investigate the stationary states of 
the system in Sec. III, in particular Bloch waves and {\color{black} spatially}
localized states. Section IV is devoted to the analytical study of the linear 
stability of the stationary states in the stack within the Bogoliubov 
approximation, and also within the hydrodynamic approach for the small coupling 
regime of Bloch waves, which is developed to a greater generality in the 
Appendix. The comparison between 
analytical and numerical results is presented in Sec. V, specially for 
transverse Josephson vortices. The nonlinear dynamics of localized states is 
discussed in Sec. VI. A final discussion together with the conclusions are 
presented in Sec. VII.


\section{Model of linearly coupled BECs}

\begin{figure}[tb]
 \centering
 \includegraphics[width=0.8\linewidth]{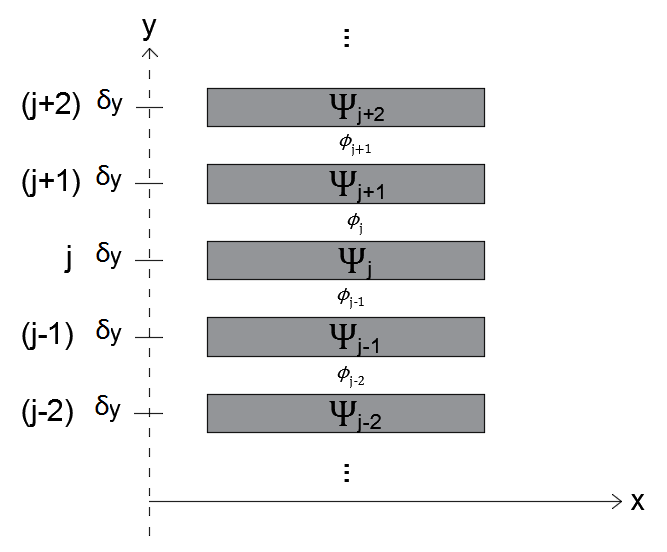}
 \caption{Sketch of a 1D-BEC array
showing the order parameters 
$\Psi_j(x,t)=\sqrt{n_j(x,t)}\exp{[i\theta_j(x,t)]}$
along each component $j$, and the relative phases 
$\phi_j(x,t)=\theta_{j+1}-\theta_{j}$ along the junctions. The discrete 
$y$-direction is effectively built through the characteristic length $\delta 
y=\sqrt{\hbar/m\Omega}$ determined by the coherent coupling of frequency 
$\Omega$ between components.}
 \label{fig:sketch}
\end{figure}
We consider a system of $M$ 1D BECs having a
coherent linear coupling of frequency $\Omega$
along the junctions.
Each BEC is described by a corresponding order parameter  
$\Psi_j(x,t)=\sqrt{n_j(x,t)}\exp{[i\theta_j(x,t)]}$,
where $n_j$ and $\theta_j$ are the local density and phase, respectively,
and $j=1,2,...M$ labels the  BECs in the stack.
We choose a particular arrangement  with periodic 
boundary conditions, so that the stack has a ring-shaped configuration with  
also $M$ coupling junctions. The $j$-th junction lies between the $j$-th and 
the $(j+1)$-th components, and the $M$-th junction connects the $M$-th and 
the $1$-st BECs. 
The total density of the system,  $n_T(x)=\sum_j\left\vert \Psi_j\right\vert 
^{2}$,
is normalized to the  total number of particles, $N= \int n_T 
\,dx = \sum_j N_j$, which is a conserved quantity,
and $N_j$ is the number of particles in the $j$-th BEC. A 
schematic representation of the system is shown in Fig.~\ref{fig:sketch}.
Detailed proposals for the experimental realization of such systems in 
ultracold gases \cite{Baals2018}, including a gauge dependent 
coupling \cite{Budich2017}, have been recently presented.

At zero temperature, within the mean-field framework, the system can be 
described by a set of coupled Gross-Pitaevskii (GP) equations, namely for  the 
$j$-th component:
\begin{align}
i\hbar\frac{\partial\Psi_j}{\partial t}  =\left( 
\frac{-\hbar^2}{2m}\partial_{x}^2 + 
g \left\vert \Psi_j\right\vert ^{2}\right)\Psi_j -
\frac{\hbar\Omega}{2} \sum_{l=j-1}^{j+1} \Psi_{l} \,,
\label{eq:Mtgp_pbc}
\end{align}
where the sum on the right hand side extends to the first
neighbours $(j-1)$ and $(j+1)$, and $m$ is the bosonic mass.
The 1D particle interaction strength is
$g=2\hbar\omega_\perp a$, the $s$-wave scattering length is 
{\color{black} assumed to be } repulsive
$a>0$, and $\omega_\perp$ is the frequency of a tight transverse trap.

For a later discussion, it is convenient to rewrite the GP 
Eq. (\ref{eq:Mtgp_pbc}) in hydrodynamic form, in terms of the densities and 
phases:
\begin{align}
\frac{\partial n_j}{\partial t}& = -\frac{\partial}{\partial x} (n_j {v}_j)
\nonumber \\\hspace{-.5cm} & -\Omega \left(
\sqrt{n_j\,n_{j+1}}\,\sin\phi_{j}-\sqrt{n_j\,n_{j-1}}\,\sin\phi_{j-1}\right) 
\,,
\label{eq:Mcontinuity}
\\ 
-\!\hbar\frac{\partial \theta_j}{\partial t} & = \mathcal{Q}_j +
\frac{m \,{v}_j^2}{2} +g n_j 
\nonumber \\ & -\frac{\hbar\Omega}{2}\left(
\sqrt{\frac{n_{j+1}}{n_j}}\,\cos\phi_{j}+\sqrt{\frac{n_{j-1}}{n_j}}\,\cos\phi_{
j-1}\right),
\label{eq:Mmomentum}
\end{align}
where {\color{black} $\phi_{j}=\theta_{j+1}-\theta_j$} is the relative phase 
between neighbor BECs, ${v}_j=\hbar\,{\partial_x}\theta_j/m $ is  
the superfluid velocity, and $\mathcal{Q}_j=-(\hbar^2 / 2m\sqrt{n_j})  
\partial_x^2 \sqrt{n_j}$ is the quantum-pressure energy term.

Equation (\ref{eq:Mcontinuity}) is the continuity equation, and expresses the 
particle conservation in the array. The local density $n_j(x)$ varies 
due to either changes in the axial current, $J_j(x)=n_j(x)v_j(x)$, within the
$j$-th BEC, or due to the Josephson current, $\mathcal{J}_{j}$, across the 
adjacent junctions. We define the latter current, by analogy with the axial 
current, as
$\mathcal{J}_{j}(x)=\mathcal{N}_{j}(x) \,\mathcal{V}_{j}(x)$, by means of a 
geometric-mean density $\mathcal{N}_j=\sqrt{n_j \,n_{j+1}}$, and a Josephson 
velocity $\mathcal{V}_{j}=({\hbar}/{m \,\delta y}) \sin\phi_{j}$, 
{\color{black} where $\delta y=\sqrt{\hbar/m\Omega}$ is the effective distance 
between BECs}. With these definitions, the last term of 
Eq.~(\ref{eq:Mcontinuity}) corresponds to a discrete derivative
$\delta \mathcal{J}_{j} / \delta y=(\mathcal{J}_{j}-\mathcal{J}_{j-1})/\delta y$
of the Josephson current. Along with the periodic boundary conditions in the 
$y$-direction, it allows for the computation of a velocity circulation around 
the stack
\begin{align}
 \Gamma=\oint v_y\,dy= \delta y \sum_{j=1}^M \mathcal{V}_{j}=
 \frac{\hbar}{m}\sum_{j=1}^M \sin\,\phi_{j} \,.
 \label{eq:ycirculation}
\end{align}
On the other hand, Eq.~(\ref{eq:Mmomentum}) is the 
equation of motion for the phase, which varies locally according to the local 
energy content. In a stationary state, where the time variation of the phase is 
given by the frequency $\mu/\hbar$ (see Eq.~(\ref{eq:BlochGen}) below), 
$\mu$ being the chemical potential, the 
local energy  in the right hand side of Eq.(\ref{eq:Mmomentum}) is the 
same (and equals $\mu$) at every point in the system.

\section{Stationary states}
\label{sec:stationary}

The BEC stack forms a discrete lattice of $M$ sites along the 
$y$-direction, transverse to the common axial $x$-axis (see 
Fig.~\ref{fig:sketch}). 
{\color{black} Since the effective (coupling-dependent) distance of separation 
between neighbor BECs is $\delta y$, the discrete coordinate along the $y$-axis 
takes values $y_j=j\delta y$
for each $j$-th BEC.} The characteristic length $\delta y$ has to be compared 
with the healing length $\xi=\hbar/\sqrt{m\,g\,n}$, determined by the axial 
density $n$ of the BECs, so that the ratio $\xi/\delta y 
=\sqrt{\hbar\Omega/gn}$ measures the tunneling-coupling strength. 

\subsection{Bloch states}
The lattice configuration along $y$ allows us 
to look for stationary states that take the form of transverse Bloch waves 
\begin{align}
\Psi_{j,k}(x,t)=\psi(x)\exp{[i(\mathcal{K}_k y_j-\mu_\mathbf{k} t/\hbar)]}\,,
\label{eq:BlochGen}
\end{align}
where $\psi(x)$ is the axial wavefunction (with the periodicity of the discrete 
lattice), and 
$\mathcal{K}_k$ is the transverse quasimomentum. Due to the discreteness of the 
system, the quasimomentum can take only $M$ different integer values within the 
first Brillouin zone:
\begin{align}
\mathcal{K}_k=\frac{2\pi}{M \delta y}\times k, \;\; \mbox{and}\;\; 
k\in\left\lbrace 
0,\,\pm 1,\,\pm 2,\dots,\floor{{M}/{2}}\right\rbrace \,,
 \label{eq:Yquasimomentum}
\end{align}
where $\floor{{M}/{2}}$ is the greatest integer less than or equal to 
$M/2$. As a result, the product space of coordinate-separated
solutions  {\color{black} (\ref{eq:BlochGen})} presents 
an $M$-fold symmetry for each wave function 
{\color{black} $\psi(x)$}. All the $k$ states but $k=0$ and $k=M/2$ (when the 
latter exists), 
at the middle and at the end, respectively, of the Brillouin zone, are states 
supporting Josephson currents in the $y$-direction, due to the existence of 
nonzero and non-$\pi$ relative phases between consecutive condensates.  

\begin{figure}[tb]
 \centering
 \includegraphics[width=0.9\linewidth]{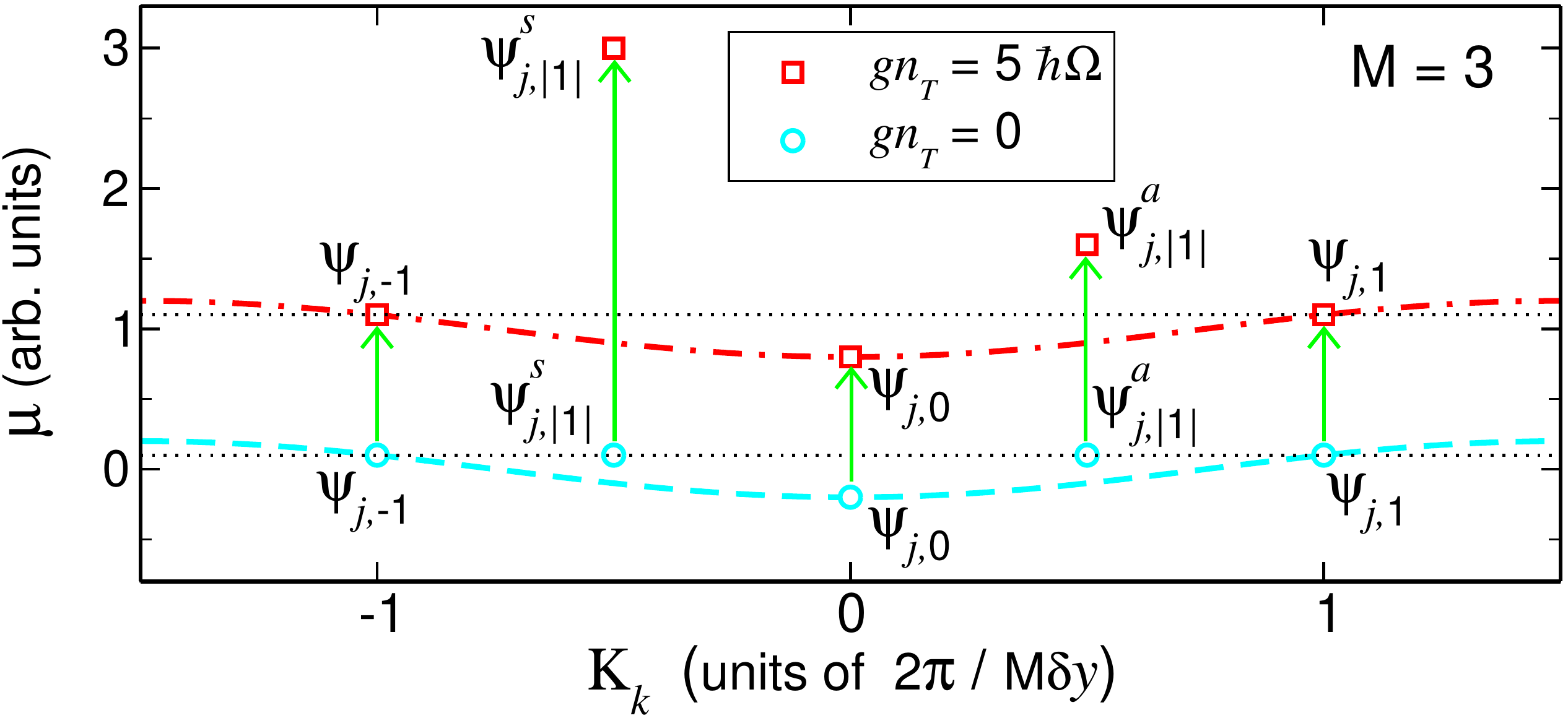}\\
 \vspace{3mm}
 \includegraphics[width=0.9\linewidth]{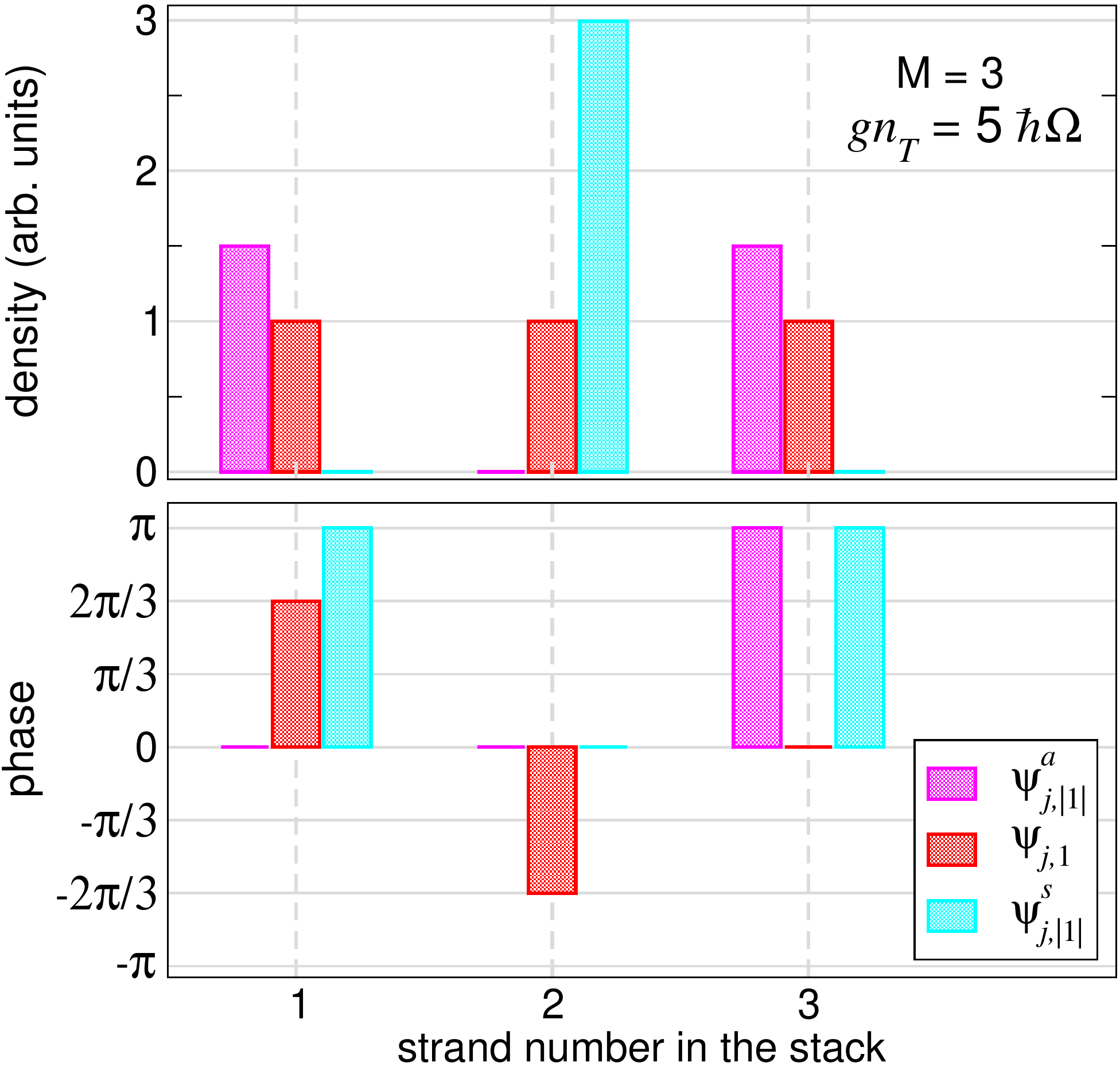}
 \caption{Stationary states in a stack made of $M=3$ coupled BECs 
with zero axial momentum, constant axial density, and 
total-interaction-to-coupling ratio  $g\,n_T/{\hbar\Omega}=5$. Top panel:  
Chemical potential of Bloch waves 
$\Psi_{j,k}$, with definite transverse quasimomentum indices $k=0,\,\pm 1$ 
as given by the horizontal axis, and standing waves $\Psi_{j,|1|}^a$ and 
$\Psi_{j,|1|}^s$, without definite quasimomentum (the horizontal axis does not 
apply) and originated at $g\,n_T/{\hbar\Omega}=0$ from linear combinations of 
Bloch waves with $|k|=1$. The vertical arrows indicate the increment in 
chemical potential obtained by the respective states from an equal increment in 
the total interaction $gn_T=5\hbar\Omega$ over the non-interacting limit. 
Bottom 
panels: Density and phase of the Bloch waves with $k=1$ (red bars) and standing 
waves with $|k|=1$ (cyan and magenta bars) in each strand $j=1,\, 2,\,3$. The 
(not represented) states $k=0,\,-1$ have, $\forall j$,
$|\Psi_{j,0}|=|\Psi_{j,-1}|=|\Psi_{j,1}|$, $\arg (\Psi_{j,0})=0$, and 
$\arg(\Psi_{j,-1})=-\arg(\Psi_{j,1})$. }
 \label{fig:M3_example_k1}
\end{figure}
In what follows we focus on states whose axial part is a momentum 
eigenstate $\psi(x)=\sqrt{n}\exp(i\mathcal{K}_x \,x)$. The 
corresponding Bloch waves read  
\begin{align}
\Psi_{j,k}(x,t)=\sqrt{n}\,\exp{[i(\mathcal{K}_x\, x+\mathcal{K}_k 
y_j-\mu_\mathbf{k} 
t/\hbar)]} \,,
\label{eq:BlochKx}
\end{align}
with stationary phases $\theta_j=\mathbf{k\cdot r}=\mathcal{K}_x 
\,x+\mathcal{K}_k y_j$, where $\mathbf{k}$ and $\mathbf{r}$ are the momentum 
and spatial vectors, respectively.
The relative phases of the Bloch waves (\ref{eq:BlochKx}) are
$\phi=\mathcal{K}_k \delta y=\pm2\pi k\,/M$, the same for all the $j$ 
junctions, and take values in the interval 
$(-\pi, \pi]$. Thus, the Josephson currents are $\mathcal{J}=\pm\hbar \,n 
\sin(2\pi k\,/M)/m \delta y$, which in the limit of large $M$ and 
small $k$ 
tend to $\mathcal{J}=n\,\hbar\,\mathcal{K}_k /m$. Analogously, the 
transverse circulation Eq.~(\ref{eq:ycirculation}) gives $\Gamma_k=\pm M\,\hbar 
\,\sin(2\pi k\,/M)/m$, and tends to $\Gamma_k=2\pi\,k\,\hbar/m$ in the same 
limit. The latter expression is equivalent to the quantized circulation of a 
regular vortex of charge $k$. {\color{black} We will refer to these discrete 
vortex currents, associated with Bloch waves having a non-zero circulation in 
the stack, as Josephson vortices}. Note that they are delocalized along 
the $x$-direction, and, unlike the regular vortices,  
do not show in general an integer-valued circulation in units of $h/m$.

By substituting the constant density states of the type (\ref{eq:BlochKx}) 
in Eq.~(\ref{eq:Mtgp_pbc}), or {\color{black} alternatively in 
Eq.~(\ref{eq:Mmomentum}), we get} the chemical potential of such Bloch waves
\begin{align}
\mu_\mathbf{k}  = g n + \frac{\hbar^2 \mathcal{K}_x^2}{2m}- \hbar\Omega 
\cos\left(\frac{2\pi k}{M}\right) \,,
\label{eq:Mmuk}
\end{align}
which takes values in the range $\mu_\mathbf{k}\in g n+ \hbar^2 
\mathcal{K}_x^2/2m+[- \hbar\Omega,\,\hbar\Omega]$, within a discrete band of 
energy width $2\hbar\Omega$. 
{\color{black} For given total density and axial momentum, the ground state of 
the system is the Bloch wave with zero quasimomentum $\Psi_{j,0}$, lying at the 
bottom of the band with $\mu_{k=0}= g 
n+\hbar^2\mathcal{K}_x^2/2m-\hbar\Omega$}. 
In the limit  $2\pi k/M \ll 1$, expanding Eq.~(\ref{eq:Mmuk}) up to quadratic 
terms one finds $\mu_\mathbf{k}  = g n-\hbar\Omega+ \hbar^2 
(\mathcal{K}_x^2 +\mathcal{K}_k^2)/2m$, which is a quadratic dispersion (as in 
a fully 2D continuous system) around the ground state.
{\color{black} The top panel of Fig. \ref{fig:M3_example_k1} depicts the 
structure of the discrete band of Bloch waves in 
the simplest case with $\mathcal{K}_x=0$ for both the non-interacting 
($gn_T=0$) and the interacting ($gn_T\neq0$) regime of 
a stack with $M=3$. In the latter, the total
interaction term ($gn_T=5\hbar\Omega$) is the same for all the 
represented states, which is equivalent to fix the whole number of particles 
$N$ in the system for given interaction strength $g$ and (finite) axial length.
The open symbols, on the top of dashed lines representing  
Eq.~(\ref{eq:Mmuk}) for a continuous index $k$, indicate the chemical potential 
of the Bloch waves $\Psi_{j,k}$ with $k=0,\,\pm 1$. The horizontal dotted lines 
serve as references for better seeing the energy degeneracies.
The two lower panels of Fig. \ref{fig:M3_example_k1} 
plot also the density and phase of the nonlinear Bloch wave with $k=1$ (red 
bars) for each strand $j=1,\,2,\,3$ (which varies along the horizontal axis). }

\subsection{Standing waves: localized states}

Along with the Bloch waves, which realize transverse current states with 
definite quasimomentum, the BEC stack admits standing-wave solutions without 
definite quasimomentum. Their existence can be tracked up to the 
non-interacting regime ($g=0$), where the standing waves can be built 
from linear combination of energy-degenerate Bloch waves (\ref{eq:BlochKx}) 
with same quasimomentum modulus $|k|$ \cite{Ringlattice2018}. Due to the 
finite size of the stack, such combinations break the lattice symmetry and show 
differences in the particle density between neighbor BECs, which lies at the 
origin of the localized-density states in periodic systems known as gap 
solitons \cite{Ringlattice2018}. It is also worth noticing that,
when integrated over the axial direction, the array of coupled 1D BECs 
can be described by a discrete nonlinear Schr\"{o}dinger model along the 
transverse $y$-coordinate, where the existence and stability of localized states
have been extensively studied \cite{Flach2008}.

Within the non-interacting regime, for each pair of energy-degenerate Bloch 
states with equal $|k|$ one can build also a pair of independent linear 
combinations with equal weight, which we 
will denote by $\Psi_{j,|k|}^s$ and $\Psi_{j,|k|}^a$, that have continuation 
into the nonlinear regime. In this way, families of nonlinear states are 
built sharing the same topology (node patterns) as the linear state from 
which they are generated. The simplest example is the $(M=3)$-stack for the 
families of states originated at $g=0$ from the (real) symmetric
$\Psi_{j,|1|}^s=\Psi_{j,1}+\Psi_{j,-1}\propto \cos(2\pi j/3)$ and  antisymmetric
$\Psi_{j,|1|}^a=\Psi_{j,1}-\Psi_{j,-1}\propto \sin(2\pi j/3)$ superpositions 
of Bloch waves with $|k|=1$. {\color{black} Since the stack has discrete 
translational symmetry along the $y$-direction, the states $\Psi_{j',|1|}^s$
and $\Psi_{j',|1|}^a$ with shifted indices $j'=j+i$ for given
$i=0,\,1,...,M-1$ (and $j' =(j+i) \pmod {M} $ if $(j+i)>M$) are degenerate 
stationary states with density 
peaks and associated phase patterns at different lattice sites.}
Figure \ref{fig:M3_example_k1} shows the chemical potential 
(top panel), and the density and the phase (at $gn_T=5\hbar\Omega$, bottom 
panels) of states belonging to these families in 
a system with zero axial momentum. Contrary to the original Bloch waves 
$\Psi_{j,\pm1}$ (either linear or nonlinear), which have equal density across 
the stack, the antisymmetric states contain a nodal strand $\Psi_{2,|1|}^a=0$ 
and two density peaks $\Psi_{1,|1|}^a=-\Psi_{3,|1|}^a$, whereas the
symmetric states present a single density peak 
$n_{2,|1|}^s>(n_{3,|1|}^s=n_{1,|1|}^s)$. As can be seen in the bottom panel 
of Fig. \ref{fig:M3_example_k1} for the symmetric state, the density 
localization increases as the states go deeper into the nonlinear regime.
For the same total density $n_T$ in the stack, {\color{black} that is for the 
same total number of particles at given axial length (as represented in the 
top panel of Fig. \ref{fig:M3_example_k1}), in the interacting case} the 
symmetric states present 
higher chemical potential than the antisymmetric ones, {\color{black} which in 
turn have higher chemical potential than the Bloch waves $\Psi_{j,\pm1}$}.

\begin{figure}[tb]
 \centering
 \includegraphics[width=\linewidth]{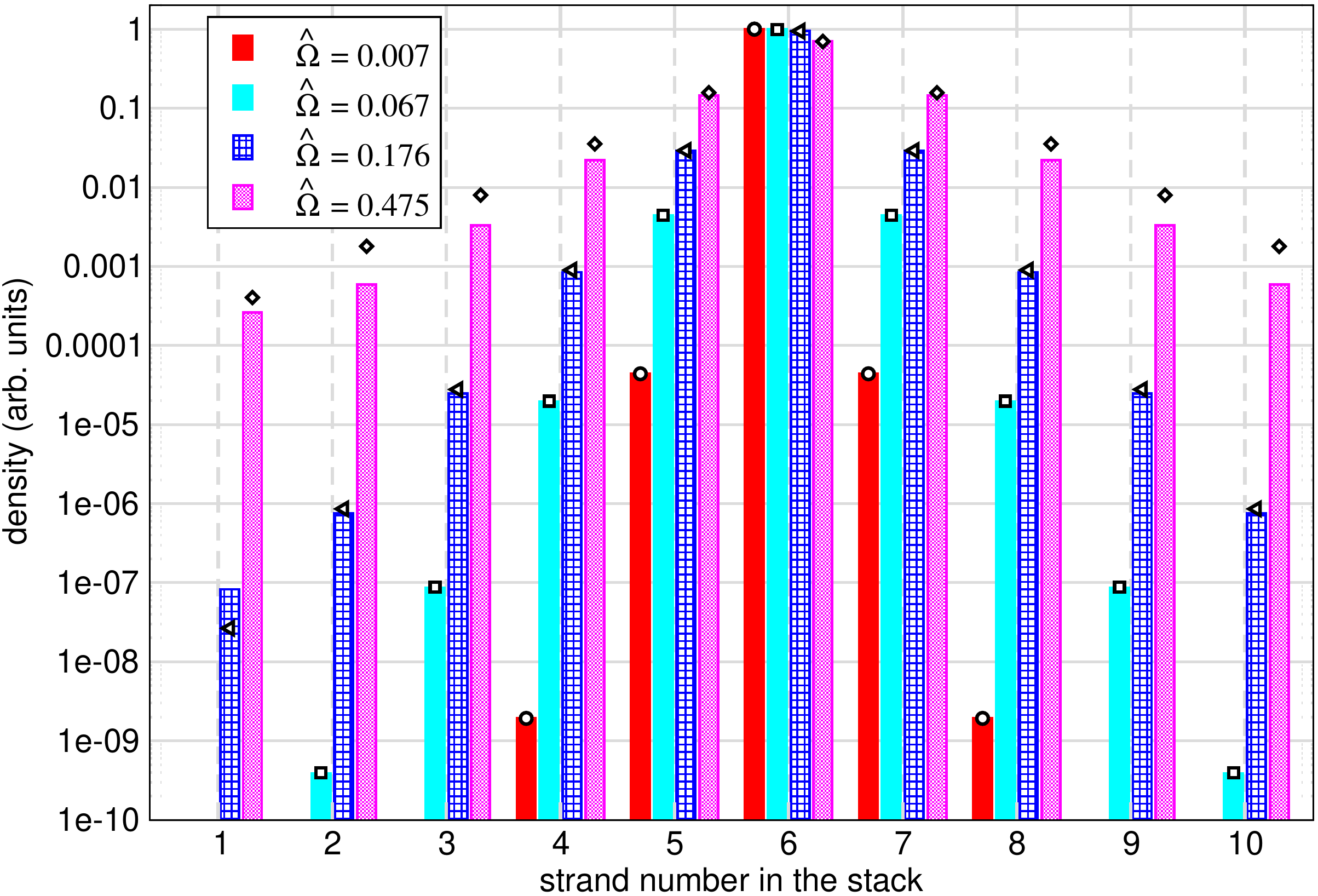}
 \caption{Stationary density profile of localized states with one density peak 
in a stack with $M=10$ and different ratios $\hat\Omega=\hbar\Omega/2gn$. The 
bars indicate the density of each BEC component as calculated from the 
numerical 
solution of the Gross-Pitaevskii equations (\ref{eq:Mtgp_pbc}), and the open 
symbols over the bars mark the corresponding value given by the approximation 
(\ref{eq:density_ratio}), derived in the small coupling regime $\hat\Omega\ll 
1$. Density values below $10^{-10}$ are not shown.}
 \label{fig:M10_localization}
\end{figure}
Although other density-localized states can be built (for example, 
{\color{black} states presenting two adjacent density peaks in either 
in-phase or out-of-phase, staggered configuration,} see e.g. 
Ref. \cite{Kivshar1993}), we will 
focus on states having either one density peak or two separated out-of-phase 
density peaks, for 
varying $M$, since they provide a representative sample for the study of 
the generic properties of localized states. As we will see {\color{black} in 
such states}, for $M>3$ and in the small coupling limit 
$\hbar\Omega/2gn \ll1$, where $n$ is now the maximum density in the stack, 
the mentioned density peaks accumulate practically the whole system density. In 
a symmetric state with $\mathcal{K}_x=0$, for instance, the nearest-neighbor 
BECs $j\pm1$ of the peak-density strand $j$ follow stationary GP equations
\begin{align}
g\,n_{j\pm1} \,\psi_{j\pm1} - \frac{\hbar\Omega}{2} (\psi_j + \psi_{j\pm2}) = 
\mu\, \psi_{j\pm1}\,,
\end{align}
which, assuming decreasing amplitudes $ \psi_j \gg \psi_{j\pm1} \gg 
\psi_{j\pm2}$, can be approximated up to first order in the neighbor amplitudes 
by $ (\hbar\Omega/2) \psi_j  \approx -\mu\, \psi_{j\pm1}$, and $\mu\approx 
g\,n$. As a result, the $l$-site BECs have amplitudes decreasing in a factor 
\begin{align}
\frac{\psi_l}{\psi_j}=\left(-\frac{\hbar\Omega}{2gn}\right)^{|l-j|} \,,
\label{eq:density_ratio}
\end{align}
for distances $|l-j|$ away from the density peak at $j$. The prototypical 
density profile is illustrated in Fig.~\ref{fig:M10_localization}, for 
different values of the coupling and same chemical potential, in a stack with 
$M=10$. The 
bars indicate the density at each $j$-th BEC as calculated from the numerical 
solution of the Gross-Pitaevskii equations, while the open symbols mark the 
values given by the analytical approximation Eq.~(\ref{eq:density_ratio}). The 
same scenario takes place for the antisymmetric states with $M>4$,
with decreasing amplitudes of opposite signs at both sides of the 
nodal strand.

\section{Linear excitations}

The almost lack of dissipation in ultracold-gas systems makes the dynamical 
stability of the stationary states the crucial issue for their experimental 
realization. For this reason, in this section we analytically address 
the linear stability of both Bloch waves and localized 
states in the stack, to be compared in next sections with their nonlinear 
dynamics. We make a general analysis based on the 
Bogoliubov equations for the linear excitations \cite{Pitaevskii2003} in order 
to find the stable regimes.
{\color{black} For Bloch waves, we have also considered a long-wavelength 
excitation approach (whose derivation is deferred to the Appendix) to derive 
a resulting system of coupled sine-Gordon-like equations similar to usual 
models in superconducting Josephson junctions \cite{Mazo2014}.}

\subsection{Linear stability of Bloch waves}

Let us consider the linear excitations $[u_{j,k}(x), v_{j,k}(x)]$ with 
energy $\hbar\,\omega$ around the 
stationary states (\ref{eq:BlochKx}). After substituting 
$\Psi_{j,k}(x,t)=\exp{(-i\mu_\mathbf{k} t/\hbar)}
[\sqrt{n}\exp{i\mathbf{(k\cdot r)}}+u_{j}(x)\exp{(-i\omega 
t)}+v^*_{j}(x)\exp{(i\omega t)}]$ in 
the GP equation (\ref{eq:Mtgp_pbc}), the excitation modes satisfy the 
Bogoliubov equations
\begin{align}
  H_\mathbf{k} \, u_j+  \, g  \, n \, e^{i2\mathbf{k\cdot r} } v_j 
-\frac{\hbar\Omega}{2} \left(u_{j-1}+u_{j+1}\right)
 & = \hbar \omega \, u_j 
\label{eq:bogu}
\\
 -H_\mathbf{k} \, v_j -  \, g  \, n \, e^{-i2\mathbf{k\cdot r}} u_j 
+\frac{\hbar\Omega}{2}  \left(v_{j-1}+v_{j+1}\right) &= \hbar \omega \, v_j \,,
\label{eq:bogv}
\end{align}
where $H_\mathbf{k} = -(\hbar^2/2m)\partial_x^2+  2 g n -\mu_\mathbf{k}$.
By making use of the Fourier expansions $ u_j(x)=\sum_{\mathbf{q}} 
c_{\mathbf{q}}\exp\{i[ (\mathcal{K}_x+q_x) x + (\mathcal{K}_k+q_p) y_j]\}$ and 
$v_j(x)=\sum_{\mathbf{q}} 
d_{\mathbf{q}}\exp\{-i[ (\mathcal{K}_x-q_x) x + (\mathcal{K}_k-q_p) y_j]\}$, 
where $q_p=2\pi p/M\delta y$ is the transverse momentum of the excitation for 
integer $p=0,\,\pm 1,\,\pm 2,\dots \floor{M/2}$, the Bogoliubov equations get 
decoupled for each two-dimensional wave number $\mathbf{q}=(q_x,q_p)$:
\begin{align}
 \left[E_{\mathbf{k},+}-\hbar\Omega \cos\left(\frac{2\pi(k+p)}{M}\right) 
\right]c_{\mathbf{q}} +  
\, g  \, n \, d_{\mathbf{q}} & = \hbar \omega \, c_{\mathbf{q}} \,,
\label{eq:bogu0}
\\
 -\left[E_{\mathbf{k},-}-\hbar\Omega \cos\left(\frac{2\pi(k-p)}{M}\right) 
\right] d_{\mathbf{q}}- \, g  \, n \, c_{\mathbf{q}}
& = \hbar \omega \, d_{\mathbf{q}} \,,
\label{eq:bogv0}
\end{align}
where $E_{\mathbf{k},\pm}= \hbar^2 (q_x^2 \pm 2\mathcal{K}_x\,q_x)/2m+  g n 
+\hbar\Omega\cos(2\pi\,k/M)$. 
Further, we introduce the linear combinations of modes $f_\pm^{(\mathbf{q})}= 
c_{\mathbf{q}}\pm d_{\mathbf{q}}$, so that
\begin{align}
\left(\zeta_{q_x} +  2 g n + \hbar\Omega\,\alpha_{k,p}
\right) f_+^{(\mathbf{q})} = 
\hbar \left(\omega-\mathcal{K}_x\frac{\hbar q_x}{m}-\Omega\,\beta_{k,p}\right) 
f_-^{(\mathbf{q})},
\label{eq:bogf+}
\\
\left(\zeta_{q_x}+ \hbar\Omega\,\alpha_{k,p}
\right) f_-^{(\mathbf{q})} =
\hbar \left(\omega-\mathcal{K}_x\frac{\hbar q_x}{m}-\Omega\,\beta_{k,p}\right) 
f_+^{(\mathbf{q})},
\label{eq:bogf_}
\end{align}
where $\zeta_{q_x}=\hbar^2 q_x^2/2m$ is the kinetic 
energy of the modes along each 1D BEC, and the transverse excitations are 
defined through the parameters
\begin{align}
 \alpha_{k,p}= \cos\left(\frac{2\pi k}{M}\right) \left[1- 
\cos\left(\frac{2\pi p}{M}\right)\right],
\end{align}
and
\begin{align}
 \beta_{k,p}= \sin\left(\frac{2\pi k}{M}\right) 
\sin\left(\frac{2\pi p}{M}\right) \,.
\end{align}
Therefore, for each stationary state $\Psi_k$, by solving Eqs. (\ref{eq:bogf+}) 
and (\ref{eq:bogf_}) for the frequency $\omega$, we get the dispersion 
relation {\color{black} of linear excitations}
\begin{align}
\hbar\omega&=\hbar\mathcal{K}_x\frac{\hbar q_x}{m}
+\hbar\Omega\,\beta_{k,p}
\nonumber \\ &
\pm\sqrt{ \left(\zeta_{q_x}+\hbar\Omega\,\alpha_{k,p}\right) 
\left(\zeta_{q_x}+\hbar\Omega\,\alpha_{k,p}+2gn\right) } \,,
 \label{eq:dispersion}
\end{align}
composed of $M$ double branches corresponding to the different values of 
$p$, {\color{black} which indexes the quasimomentum excitation.
Equation ~(\ref{eq:dispersion}) provides the general result, as a 
function of the parameters of the system $\{M,\,\Omega,\, gn\}$, 
for the linear stability of Bloch states with momentum 
$\mathbf{k}=(\mathcal{K}_x,\mathcal{K}_k)$ in a stack of coupled BECs.
  It gives rise to complex frequencies, associated to dynamical instabilities, 
for negative values of the 
expression inside the square root. These are modulational instabilities 
that break down the homogeneous density profile across the stack. They can only 
appear if $\alpha_{k,p}$ takes 
negative values, which occurs for $k>M/4$. Hence, all the Bloch states with 
constant density are dynamically stable if $k\leq M/4$, irrespective of the 
coupling $\Omega$. For increasing $M$, the first stable Josephson current 
($k\neq 0$)  states correspond to $M=4$ and $k=\pm 1$, which are discrete 
transverse vortices with circulation $\Gamma_{\pm 1}= \pm 4\hbar/m$ (see Sect. 
\ref{sec:bloch} below). This quasimomentum-dependence stability of the 
Bloch waves resembles similar features of the discrete nonlinear Schr\"odinger 
equation \cite{Kivshar1993,Kivshar1992}. {\color{black} As we show in Sects. 
\ref{sec:bloch} and \ref{sec:localized} by analyzing some particular 
examples, the stability features of Bloch states predicted by 
Eq.~(\ref{eq:dispersion}) are confirmed by the nonlinear dynamics as 
obtained from the numerical simulations of the time-dependent GP 
Eq.~(\ref{eq:Mtgp_pbc}).}

Among the dispersion branches of Eq.~(\ref{eq:dispersion}), the $p=0$ branch is 
always gapless (because $\alpha_{k,0}=\beta_{k,0}=0$), whereas the rest 
present energy gaps given {\color{black}(at $q_x=0$)} by 
\begin{align}
\hbar \omega_g=  \hbar\Omega\,\beta_{k,p}\pm\sqrt{ 
\hbar\Omega\,\alpha_{k,p}\left(\hbar\Omega\,\alpha_{k,p} + 2g\,n\right)} \,,
 \label{eq:gap}
\end{align}
which show up even in the non-interacting ($g=0$) case.
The speed of sound along each BEC (i.e. along the $x$ coordinate) can be 
calculated from the gapless branch {\color{black} of the dispersion} in the 
long-wavelength limit. In the frame of reference moving with axial velocity 
$\hbar\mathcal{K}_x/m$, it is:
\begin{align}
c=\left(\frac{\partial \omega}{\partial q_x}\right)_{q_x\rightarrow 0}= 
\sqrt{\frac{gn}{m}} \,.
\label{eq:sound}
\end{align}
These quantities, $\hbar\omega_g$ and $c$, are relevant for the energetic 
stability of the system, since they define thresholds, in energy  and speed, 
respectively, for the superfluid excitation by external perturbations.

In order to solve for the excitation spectrum, we choose 
the usual Bogoliubov normalization,  $\int dx \,(|u_j|^2-|v_j|^2)=1$, for the
modes $(u_j,v_j)$ that have real energy values. By selecting also real values 
for the Fourier amplitudes $c_{\mathbf{q}}$ and $d_{\mathbf{q}}$, it follows 
that 
$|c_{\mathbf{q}}|^2-|d_{\mathbf{q}}|^2=f_+^{(\mathbf{q})}\,f_-^{(\mathbf{q}) }
=1/L_x$, where $L_x$ is the axial length of the BECs. With this prescription we 
solve Eqs.~(\ref{eq:bogf+})-(\ref{eq:bogf_}) for the {\color{black} 
stable} excitation modes
\begin{align}
f_+^{(\mathbf{q})}=\pm\frac{1}{\sqrt{L_x}}\left( 
\frac{\zeta_{q_x}+\hbar\Omega\,\alpha_{k,p}}
{\zeta_{q_x}+\hbar\Omega\,\alpha_{k,p}+2gn} \right)^{1/4} \,.
 \label{eq:fmodes_real}
\end{align}

{\color{black} The unstable excitation modes are associated with the complex 
energies of Eq.~(\ref{eq:dispersion}).} The corresponding 
normalization reads $\int dx \,(|u_j|^2-|v_j|^2)=0$ \cite{Pitaevskii2003}, and 
we set $\int dx \,|u_j|^2=1$. Then 
$d_{\mathbf{q}}=c_{\mathbf{q}}\,\exp(i2\varphi_{\mathbf{q}})$, and the unstable
excitation modes are
\begin{align}
f_\pm^{(\mathbf{q})}=\frac{1}{\sqrt{L_x}}[1\pm 
\exp(i2\varphi_{\mathbf{q}})]\,,
\end{align}
where, from Eqs.~(\ref{eq:bogf+})-(\ref{eq:bogf_}), the phase 
$\varphi_{\mathbf{q}}$ is given by
\begin{align}
\varphi_{\mathbf{q}}=\mathrm{atan}\left(\pm\sqrt{ 
-\frac{\zeta_{q_x}+\hbar\Omega\,\alpha_{k,p}+2gn}{\zeta_{q_x}+\hbar\Omega\,
\alpha_{k,p}} }\,\right).
 \label{eq:fmodes_complex}
\end{align}

\subsubsection{Limit of long-wavelength excitations and small coupling.}

As follows from the hydrodynamic approach for the linear 
excitations of Bloch waves developed in the Appendix, in the 
long-wavelength regime and small Josephson coupling, 
$(\hat\Omega=\hbar\Omega/2gn)\ll 1$, the equations of motion for excitations in 
the relative phase $\phi_{j}$ and relative density $\rho_{j}$ become:
\begin{align}
\frac{1}{c^2}\frac{\partial^2 \rho_{j}}{\partial t^2}-\frac{\partial^2 
\rho_{j}}{\partial x^2}-\alpha_k \frac{\delta^2\rho_{j}}{\delta y^2} = 0 \,,
\label{eq:dens_wave}
\\
\frac{1}{c^2}\frac{\partial^2 \phi_{j}}{\partial t^2}-\frac{\partial^2 
\phi_{j}}{\partial x^2} -\alpha_k \frac{\delta^2\sin\phi_{j}}{\delta y^2} \!= 
\!
\beta_k\frac{\rho_{j+1}-\rho_{j-1}}{\delta y^2} \,,
\label{eq:phase_wave}
\end{align}
where $\alpha_k=\cos(2\pi\,k/M)$ and $\beta_k=\sin(2\pi\,k/M)$.
This system of $M$ pairs of equations describes the linear dynamics of the 
underlying array of junctions determined by the 
GP Eqs.~(\ref{eq:Mtgp_pbc}). The $k$-dependent factor $\alpha_k$ 
multiplying the transversal (discrete) derivative indicates a varying 
penetration length $\xi_{J,k}=\delta y /\sqrt{|\alpha_k|}$. The sine functions 
in Eq.~(\ref{eq:phase_wave}) should be formally 
substituted by their arguments $\sin \phi_j \approx \phi_j$, since we have 
assumed a 
linear regime. By keeping them, we highlight the correspondence with a 
sine-Gordon-like equation, where kink solutions can be found \cite{Barone1982}.
As a limiting case, the kink-type solutions have demonstrated to be useful in 
the search of solitonic states in the nonlinear dynamics of two-component 
condensates \cite{Qadir2012a,Sophie2018,Son2002,Qu2017,Su2015}. 

Equations (\ref{eq:dens_wave})-(\ref{eq:phase_wave}) admit plane wave 
solutions with the same phase shifts across the stack previously found in 
Eq.~(\ref{eq:Yquasimomentum}):
\begin{align}
\rho_{j,k,p}(x,t)&=& c_{\mathbf{q}}\, \exp{[i(q_x x + q_p y_j -\omega_{k,p} 
\,t)]} \,,
\nonumber\\
\phi_{j,k,p}(x,t) &= & d_{\mathbf{q}}\, \exp{[i(q_x x + q_p y_j -\omega_{k,p}\, 
t)]} \,,
\label{eq:waves}
\end{align}
with transverse momentum $q_p=2\pi p/M\delta y$. After substitution, 
one gets the amplitude relation $c_{\mathbf{q}}= i \,\nu \,d_{\mathbf{q}}$, 
where $\nu$ 
is a real number, and the double-branched dispersion 
\begin{align}
\omega_{k,p}^{(1)}=c \, \sqrt{\,q_x^2+\frac{2}{\delta y^2}\,\alpha_{k,p}} 
\;\;,
\nonumber\\
\omega_{k,p}^{(2)}=c \, \sqrt{\,q_x^2+\frac{2}{\delta 
y^2}\,(\alpha_{k,p}+\nu\beta_{k,p})}\;\;.
\label{eq:hydro_dispersion}
\end{align}
For $\beta_k=0$ (that is, when $k=0$ or $k=M/2$, and $\alpha_k=\pm 1$) both 
branches coincide, in agreement with the Bogoliubov dispersion 
Eq.~(\ref{eq:dispersion}) 
in the small coupling limit, $\hat\Omega\ll 1$, considered 
here. The gaps are 
given by $\omega_{k,p}^{(1)}=c\sqrt{2\alpha_{k,p}}/\delta y$.
Additionally, for $q_p \ll 1$ the dispersion can be written as
$\omega_{k,p}=c \,\sqrt{q_x^2 \pm  q_p^2}$, which is the 
usual dispersion $\omega=c \,|\mathbf{q}| $ of a continuous (relativistic) 2D 
system when $k=0$, and contains instabilities when $k=M/2$ for 
$|q_p|>|q_x|$.

The second dispersion branch $\omega_{k,p}^{(2)}$, is not fully consistent 
with the Bogoliubov spectrum for $\beta_{k,p}\neq 0$, unless 
$\nu = 1$ and the limit $\hbar\Omega\rightarrow 0$ were considered. Such
inconsistencies have been previously found in two coupled condensates 
(see e.g. \cite{Abad2013}), and arise from neglecting the quantum pressure 
term. {\color{black}As a consequence, and as usual within this approximation, 
the validity of this hydrodynamic model gets restricted to the lowest energy 
excitations (long-wavelength excitation modes of the lowest energy branch), 
which is equivalent to consider decoupled wave equations 
(\ref{eq:dens_wave})and (\ref{eq:phase_wave}), neglecting the right hand side 
of the latter, for the relative densities and phases.}

\subsection{Linear stability of localized states}
\label{sec:localized_stab}
We study the stability of {\color{black} the previously introduced} 
standing-wave 
states having one or two density peaks in the stack of constant density BECs. 
The analysis of the 
simplest systems with $M=3,\, 4$ components serves as an insightful starting 
point, with straightforward analytical solutions for the antisymmetric states. 
For a generic $M$ value, we focus on the small coupling limit 
$(\hat\Omega=\hbar\Omega/2gn) \ll 1$, where now $n$ is the peak (or maximum) 
density in the stack.

The wavefunctions of the antisymmetric states with $M=3$ are: 
$\Psi_{2,|1|}^a=0$ 
and $\Psi_{3,|1|}^a=-\Psi_{1,|1|}^a=\sqrt{n}\,\exp[i(\mathcal{K}_x\, 
x-\mu_{|1|}^a t/\hbar)]$, where the chemical potential is 
$\mu_{|1|}^a=\hbar^2\mathcal{K}_x^2/2m+gn+\hbar\Omega/2$. 
Note that the stationary phase pattern of these nonlinear states (for 
$\mathcal{K}_x=0$ and $t=0$) is determined by the corresponding 
linear state of this family (at $g=0$), {\color{black} in this case we set 
$\Psi_{j,|1|}^a \propto \sin(2\pi\,(j+1)/3)$}. Two pairs of Bogoliubov 
equations are obtained, namely for the first BEC with $j=1$: 
\begin{align}
  H_{|1|}^{a} \, u_1+  \, g  \, n \, e^{i2\mathcal{K}_x x } v_1
-\frac{\hbar\Omega}{2} \left(u_{2}+u_{3}\right) & = \hbar \omega \, u_1 \, ,
\label{eq:abogu1}
\\
 -H_{|1|}^{a} \, v_1 -  \, g  \, n \, e^{-i2\mathcal{K}_x x} u_1 
+\frac{\hbar\Omega}{2}  \left(v_{2}+v_{3}\right) &= \hbar \omega \, v_1 \,,
\label{eq:abogv1}
\end{align}
where $H_{|1|}^a=-(\hbar^2/2m)\partial_{x}^2 +  2g n-\mu_{|1|}^a$, and 
identical equations follow for the $j=3$ BEC component by swapping indexes $1$ 
and $3$; and for the nodal $j=2$ component 
\begin{align}
  \left(-\frac{\hbar^2}{2m}\frac{\partial^2}{\partial x^2}-\mu_{|1|}^a\right)  
u_2-\frac{\hbar\Omega}{2} \left(u_{1}+u_{3}\right) & = \hbar \omega \, u_2 \, , 
\label{eq:abogu2}
\\
 -\left(-\frac{\hbar^2}{2m}\frac{\partial^2}{\partial x^2}-\mu_{|1|}^a\right) 
v_2 
+\frac{\hbar\Omega}{2}  \left(v_{1}+v_{3}\right) &= \hbar \omega 
\, v_2 \,.
\label{eq:abogv2}
\end{align}
As previously, making use of the Fourier expansions
$ u_j(x)=\sum_{q} c_{q}\exp\{i[ (\mathcal{K}_x+q_x) x]\}$ and 
$v_j(x)=\sum_{q} d_{q}\exp\{-i[(\mathcal{K}_x-q_x) x]\}$, after some 
algebraic manipulations, we get the following six branches of the spectrum:
\begin{align}
\hspace{-1cm}
&\hbar\omega_{|1|}^a  = \;\hbar\mathcal{K}_x\frac{\hbar q_x}{m}
\pm\sqrt{(\zeta_{q_x}-{\hbar\Omega})(\zeta_{q_x}-{\hbar\Omega}+2gn)},
\nonumber\\
\hspace{-1cm}
&\hbar\omega_{|1|}^a = \;\hbar\mathcal{K}_x\frac{\hbar q_x}{m} \nonumber \\  &
\pm\sqrt{\left(\zeta_{q_x}-\frac{3\hbar\Omega}{2}\right)\left(\zeta_{q_x}-\frac{
3\hbar\Omega}{2} \pm 
2gn\gamma^{(3)}\right)-\lambda^{(3)}\frac{\hbar\Omega}{2}gn 
},
\label{eq:adispersion}
\end{align}
where $\gamma^{(3)}=\sqrt{(1-\hat\Omega/2)^2+2\hat\Omega^2}$, and 
$\lambda^{(3)}=1-9\hat\Omega/4$. As is apparent {\color{black} from the 
negative values of the radicand}, the system is unstable.

The analysis is simpler for the antisymmetric states with $M=4$, where
the wave functions are 
$\Psi_{3,|1|}^a=-\Psi_{1,|1|}^a=\sqrt{n}\,\exp[i(\mathcal{K}_x\, x-\mu_{|1|}^a 
t/\hbar)]$, and $\Psi_{2,|1|}^a=\Psi_{4,|1|}^a=0$, and the chemical potential 
does not depend on the coupling $\mu_{|1|}^a=\hbar^2\mathcal{K}_x^2/2m+gn$. 
Similar manipulations of the corresponding Bogoliubov equations as before, 
which coincide 
with Eqs.~(\ref{eq:abogu1}) to (\ref{eq:abogv2}), lead to the eight branches of 
the spectrum
\begin{align}
\hbar\omega_{|1|}^a&= \;\hbar\mathcal{K}_x\frac{\hbar q_x}{m} \pm 
(\zeta_{q_x}-gn) \,,
\nonumber\\
\hbar\omega_{|1|}^a&= \;\hbar\mathcal{K}_x\frac{\hbar q_x}{m}
\pm\sqrt{\zeta_{q_x}(\zeta_{q_x}+2gn)} \,,
\label{eq:a4dispersion}\\
\hbar\omega_{|1|}^a&= \;\hbar\mathcal{K}_x\frac{\hbar q_x}{m}
\pm\left(\gamma^{(4)}\pm\sqrt{(\gamma^{(4)})^2-(\lambda^{(4)})^2} 
\right)^{\frac{1}{2}} \,,
\nonumber
\end{align}
where $\gamma^{(4)}=\zeta_{q_x}^2+(\hbar\Omega)^2+(gn/2)^2$ and 
$\lambda^{(4)}=[\zeta_{q_x}^2-(\hbar\Omega)^2-(gn)^2\,]^2-(gn)^2(\zeta_{q_x}
-gn)^2$. Different to the $M=3$ case, and although the last expression in 
(\ref{eq:a4dispersion}) can in general produce imaginary frequencies, it is 
still possible to find dynamically stable states (with only real frequencies) 
in the small coupling regime whenever  $\lambda^{(4)}<\gamma^{(4)}$.

The symmetric states involve a more cumbersome algebra. For the simplest stack 
with $M=3$, the chemical potential 
is $\mu_{|1|}^s=\hbar^2\mathcal{K}_x^2/2m+(2gn_{1,|1|}^s+g\delta 
n_{1,|1|}^s-\hbar\Omega/2+\sqrt{(g\delta 
n_{1,|1|}^s+\hbar\Omega/2)^2+2(\hbar\Omega)^2\,})/2$, where $\delta n_{1,|1|}^s=
n_{2,|1|}^s-n_{1,|1|}^s$ is the density contrast between components, which 
increases with the chemical potential for a given coupling. An asymptotic 
analysis can be readily done in the small coupling limit $\hat \Omega\ll 1$, 
where the densities fulfill $(n_{2,|1|}^s\approx n_T) \gg 
(n_{1,|1|}^s=n_{3,|1|}^s)$. 
Then $\Psi_{2,|1|}^s=\sqrt{n}\exp[i(\mathcal{K}_x\, x-\mu_{|1|}^s t/\hbar)]$ and
$(\Psi_{1,|1|}^s=\Psi_{3,|1|}^s)\approx -\hat \Omega\,\Psi_{2,|1|}^s$, and the 
chemical potential is $\mu_{|1|}^s\approx \hbar^2\mathcal{K}_x^2/2m+gn$. In 
this approximation, the Fourier expansion of the excitation modes leads to the 
same functional form of the spectrum (\ref{eq:a4dispersion}) with the 
substitution of $\hbar\Omega$ by $\hbar\Omega/\sqrt{2}$. As a consequence, 
dynamical stability of the (one-peak) symmetric states is also expected in the 
small $\hat\Omega$ regime. In this case, the 
strong localization of the density in a single strand, with neighbor 
densities decreasing as powers of $\hat\Omega^2$, allows the stability 
prediction to be extended to arbitrary $M$, and even to systems with open 
boundary conditions. Furthermore, the same analysis also applies to 
the (two-peak) antisymmetric states with $M>4$. As we later demonstrate in 
Sect. \ref{sec:localized}, the numerical solutions of both the Bogoliubov 
equations for the linear excitations and the GP equation for the nonlinear time 
evolution confirm these predictions.

\section{Transverse Josephson vortices}
\label{sec:bloch}
\begin{figure}[tb]
 \centering
  \includegraphics[width=0.9\linewidth]{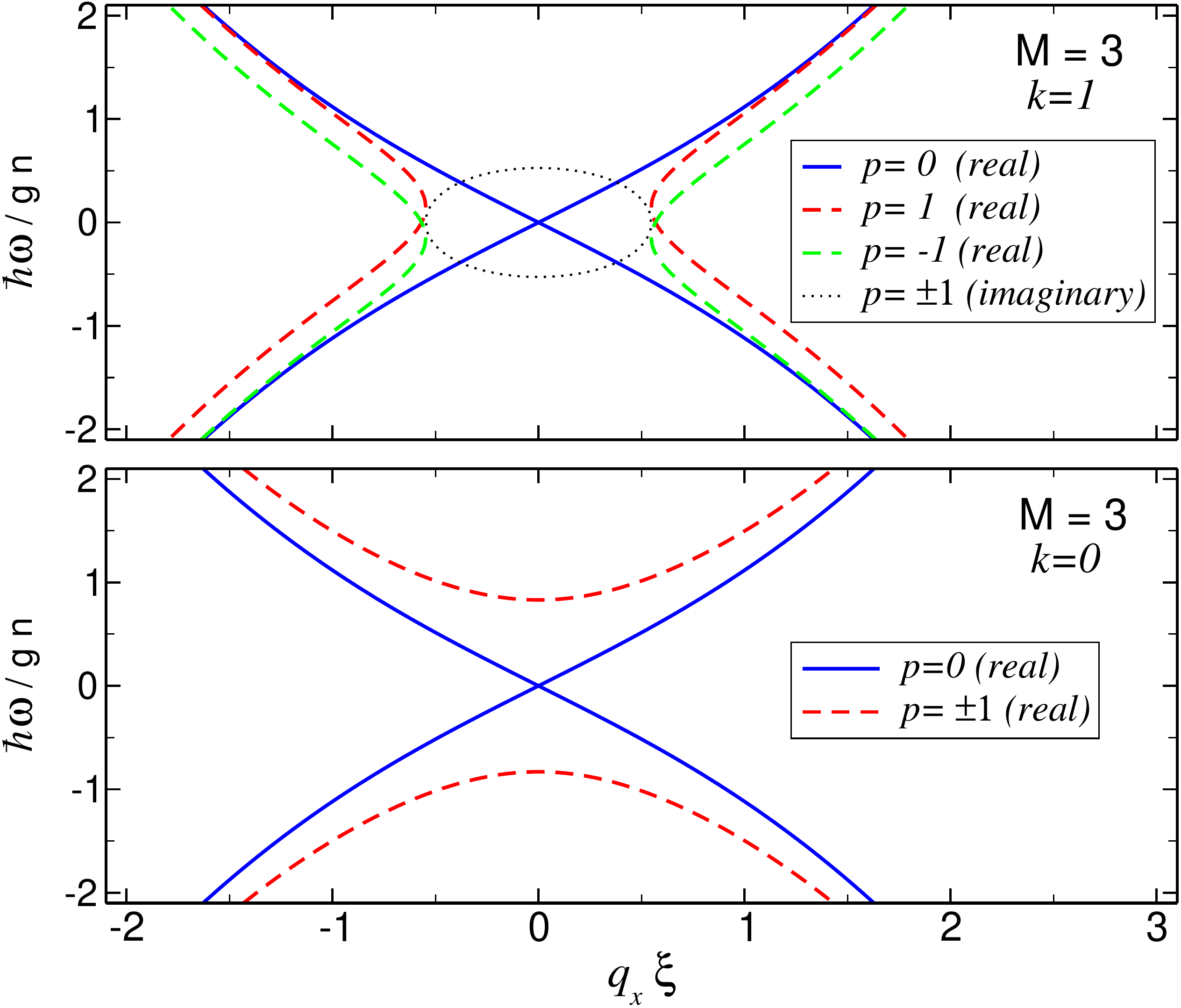}
 \caption{Dispersion relations for constant density states in a $M=3$ stack 
with $\hbar \Omega=0.2 \, gn$ and $\mathcal{K}_x=0$. {\color{black}The 
$p$ index labels the excitation modes of each stationary state (see text). The 
horizontal axis measures the axial quasimomentum excitation $q_x$ in (inverse 
of) healing length $\xi$ units. Top panel: Energy excitations of 
the Bloch wave with $k=1$ and chemical potential $\mu_1/gn=1.1$. For the 
unstable modes $p=\pm 1$, only the imaginary part of 
the complex frequency $\hbar\,$Im($\omega$) is represented (dotted line). 
Bottom panel: Energy excitations of the ground state with $k=0$ and 
$\mu_0/gn=0.8$.}}
 \label{fig:M3disp}
\end{figure}
\begin{figure}[tb]
 \centering
  \includegraphics[width=0.9\linewidth]{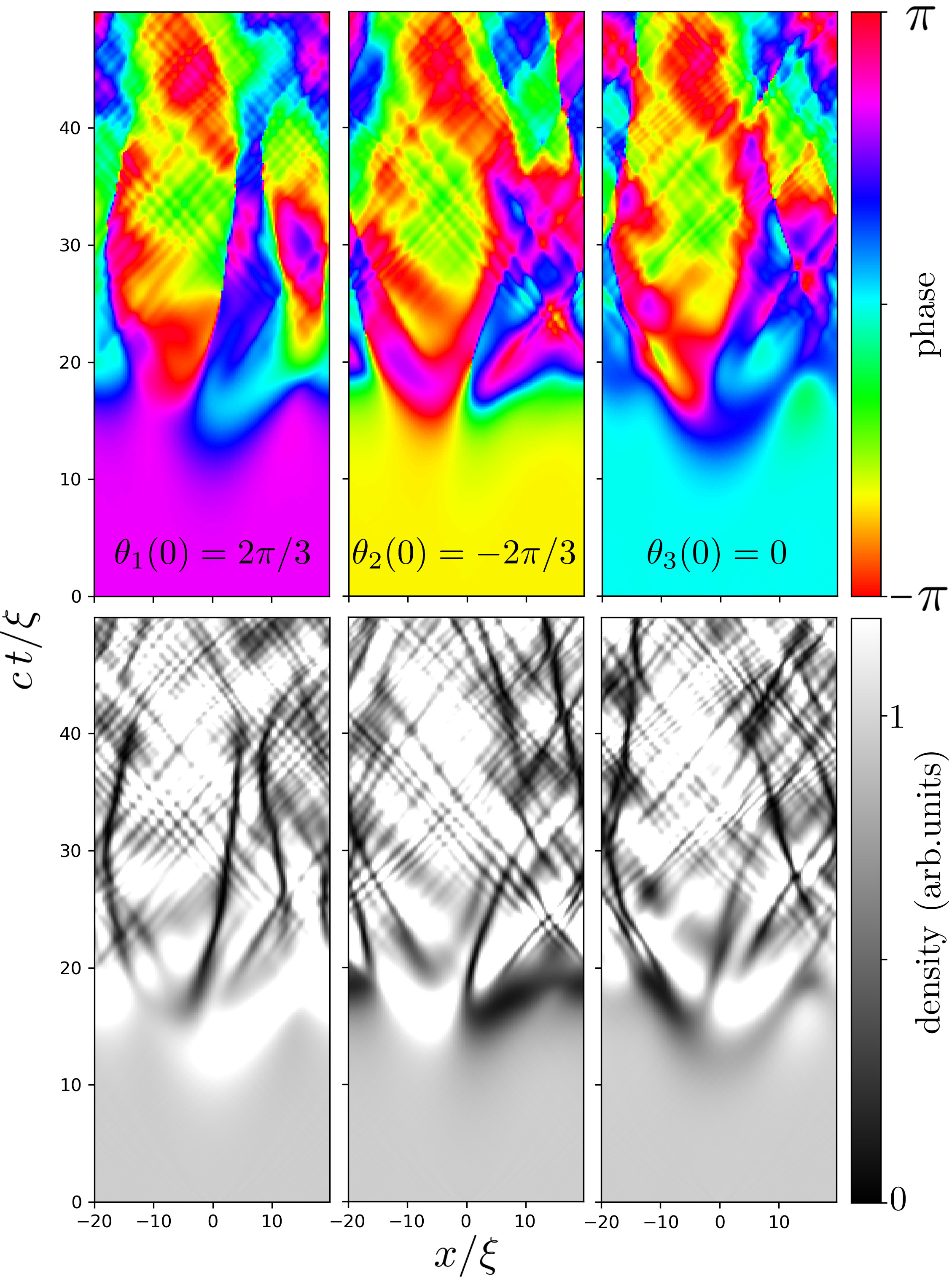}
 \caption{Decay of a Bloch state with $k=1$ and constant density in 
a $M=3$ stack with $\hbar \Omega=0.2 \, gn$ and $\mathcal{K}_x=0$. The time 
evolution of the axial phases (top panels) and the axial densities (bottom 
panels) is shown for each BEC component, from left to right, $j=$1, 2, 3.
The labels in the top panels indicate the corresponding phases
$\theta_j$ of the stationary configuration. The initial state  (at $t=0$) has 
been seeded with a random Gaussian perturbation (less than 1$\%$ in 
amplitude). {\color{black} On the horizontal axis, the axial length is measured 
in healing length $\xi$ units.}}
 \label{fig:evolM3_k1}
\end{figure}
In this section we compare the stability predictions of the linear analysis 
{\color{black} for Bloch wave states} with the numerical solutions of the GP 
Eq. 
(\ref{eq:Mtgp_pbc}) for representative sets of parameters. Just for the 
purpose of showing two different examples, we consider one case with zero axial 
momentum $\mathcal{K}_x=0$ and another with $\mathcal{K}_x \neq 0$. As we will 
see, the linear predictions coincide with the numerical results of the 
nonlinear dynamics, and in particular, stable transverse Josephson vortices 
{\color{black} (associated with Bloch waves having non-zero circulation)} are 
found.

\subsection{Case $M=3$ and $\mathcal{K}_x=0$}
\label{sec:blochA}

The Bloch waves have quasimomentum $\mathcal{K}_k$ for values of  $k=0,\pm 1$. 
Let us analyze the solutions for each particular case:
\begin{itemize}
 \item $k=0 \,$: {\color{black} This is the ground state for given interaction 
strength and total density.} All the components share the same wave 
function 
$\psi_{0}=\sqrt{n}\,\exp{(-i\mu_0 t/\hbar)}$, with
chemical potential  $\mu_0=gn-\hbar\Omega$. By using 
the parameters $\alpha_{0,p}=\{0\,,3/2,\,3/2\}$ and $\beta_{0,p}=\lbrace 
0,0,0\rbrace$, with $p=0,\pm 1$, the dispersion curves, $\omega_{0,p}\,$, are
\begin{align}
\hbar \omega_{0,0}&=&\pm\sqrt{ \zeta_{q_x} \left(\zeta_{q_x}+2gn\right)} \,, \\
\hbar\omega_{0,\pm1}&=&\sqrt{\left(\zeta_{q_x}+\frac{3\hbar\Omega}{2}
\right)\left(\zeta_ { q_x}+\frac{3\hbar\Omega}{2}+2gn\right)} \,.
 \label{eq:dispersionM3_0}
\end{align}
These expressions are plotted in the bottom panel of Fig.~\ref{fig:M3disp} for 
a system with $\hbar\Omega=0.2 \, gn$. {\color{black} As expected, all the 
excitation energies are real, and the state is stable.}

\item $k=1\,$: The wave functions for the BEC components are:  
\begin{align}
\psi_{1,1}&=\sqrt{n}\,\exp{[ i(2\pi/3-\mu_1 t/\hbar)]} \,,\\ 
\psi_{2,1}&=\sqrt{n}\,\exp{[ i(4\pi/3-\mu_1 t/\hbar)]} \,,\\ 
\psi_{3,1}&=\sqrt{n}\,\exp{[ i(2\pi-\mu_1 t/\hbar)]} \,,
\end{align}
which configure a discrete anti-vortex of circulation $\Gamma_1= 
- 3\sqrt{3}\hbar/2m$ around the discrete $y$-direction. 
Analogously, the Bloch wave with $k=-1$ corresponds to an (energetically 
degenerate) vortex with opposite circulation 
$\Gamma_{-1}= -\Gamma_1$. The chemical potential is 
$\mu_1=gn+\hbar\Omega/2$ and the 
parameters $\alpha_{1,p}=\{0\,,-3/4,\,-3/4\}$ and 
$\beta_{1,p}=\{0\,,3/4,\,-3/4\}$. 
The resulting dispersion curves, $\omega_{1,p} \,$, are:
\begin{align}
\hbar \omega_{1,0}&=\pm\sqrt{ \zeta_{q_x} \left(\zeta_{q_x}+2gn\right)} \,, \\
\hbar\omega_{1,1}&=\frac{3\hbar\Omega}{4}
\\ \nonumber &
\pm \sqrt{\left(\zeta_{q_x}-\frac{
3\hbar\Omega}{4}\right)\left(\zeta_ { q_x}-\frac{3\hbar\Omega}{4}+2gn\right)} 
\,,\\
\hbar\omega_{1,-1}&= -\frac{3\hbar\Omega}{4} 
\\ \nonumber &
\pm \sqrt{\left(\zeta_{q_x}-\frac{3\hbar\Omega}{4}
\right)\left(\zeta_ {q_x}-\frac{3\hbar\Omega}{4}+2gn\right)} \,.
 \label{eq:dispersionM3_1}
\end{align}
The upper panel of Fig.~\ref{fig:M3disp} depicts these expressions for a 
system with $\hbar\Omega=0.2 \, gn$. For $\omega_{1,\pm1}\,$, the negative 
signs under the square root indicate the presence of instabilities on the 
vortex state.
An example of the decay dynamics of this unstable vortex state is shown in 
Fig.~\ref{fig:evolM3_k1}. The graph depicts the real time evolution of the 
system after imprinting a small random perturbation on the stationary state. 
The data, axial phases (top panels) and axial densities (bottom panels) for 
each component, have been obtained from the numerical solution of the GP Eqs. 
(\ref{eq:Mtgp_pbc}) with periodic boundary conditions in the axial coordinate, 
given in units of the healing length $\xi$. As can be seen, the stationary 
configuration survives for a time lapse of around $t\approx 10 \,\xi/c$, beyond 
which soliton-like structures (tracing thick, dark paths on the axial density 
plots) appear, producing strong density and phase modulations. {\color{black} 
Different noise seeds on the initial state produce different density and phase 
patterns during the decay dynamics, with the only common feature of the 
emergence of several, interacting solitons.}
\end{itemize}

\begin{figure}[tb]
 \centering
  \includegraphics[width=0.9\linewidth]{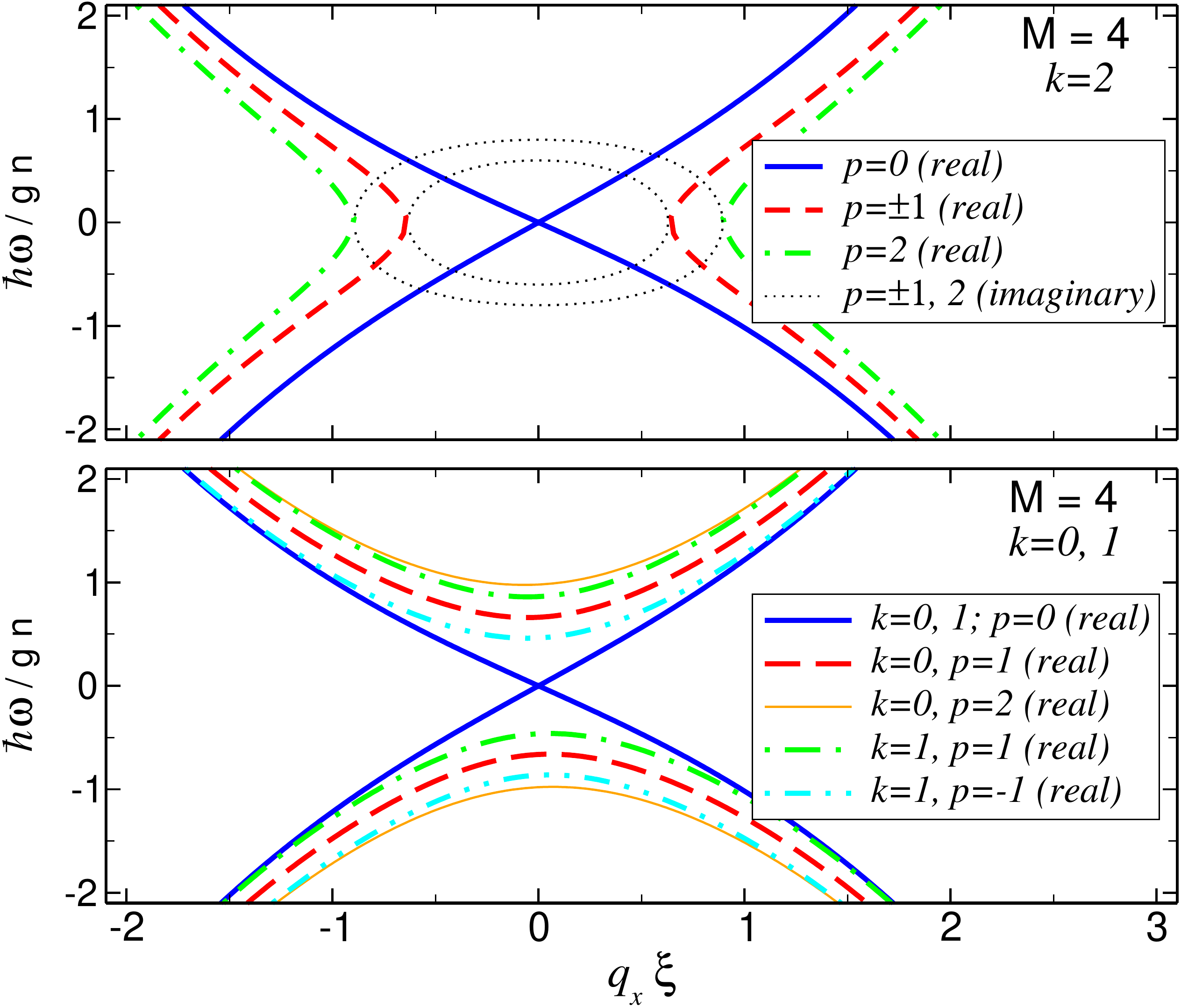}
 \caption{{\color{black}Same as Fig. \ref{fig:M3disp}} for constant density 
states in a $M=4$ stack with $\hbar \Omega=0.2 \, gn$ and $\mathcal{K}_x 
\,\xi=0.1$. The top panel corresponds to the (unstable) Bloch wave with $k=2$, 
whereas the bottom panel shows the linear excitations of the (stable) Bloch 
waves with $k=0$ and $k=1$. Only the imaginary part (dotted curves) is plotted 
for the unstable frequencies.}
 \label{fig:M4disp}
\end{figure}

\subsection{Case $M=4$ and $\mathcal{K}_x\neq0$: stable Josephson 
{\color{black} vortices with $k=\pm1$} }
\label{sec:blochB}

Here we consider arbitrary axial momentum ($\mathcal{K}_x$) states. The 
Bloch waves have $k=0,\pm 1,\, 2$, and the excitation modes have also  
$p=0,\pm1,2$.
\begin{itemize}
\item $k=0 \,$: This corresponds to the ground state for
a given $\mathcal{K}_x$, 
$\psi_{0}=\sqrt{n}\,\exp{(i\mathcal{K}_x\,x-i\mu_0 t/\hbar)}$, with
chemical potential
$\mu_0=gn+{\hbar^2 \mathcal{K}_x^2}/{2m}-\hbar\Omega$, and 
system parameters $\alpha_{0,p}=\{0\,,1,\,1,\,2\}$ and 
$\beta_{0,p}=\lbrace 0,0,0,0\rbrace$. The dispersion curves read:
\begin{align}
\hbar \omega_{0,0}&=\frac{\hbar^2 
\mathcal{K}_x q_x}{m}\pm\sqrt{ \zeta_{q_x} 
\left(\zeta_{q_x}+2gn\right)} \,, \\
\hbar\omega_{0,\pm1}&=\frac{\hbar^2 
\mathcal{K}_x q_x}{m}
\\ \nonumber &
\pm\sqrt{\left(\zeta_{q_x}+{\hbar\Omega}
\right)\left(\zeta_ { q_x}+{\hbar\Omega}+2gn\right)} \,,\\
\hbar\omega_{0,2}&=\frac{\hbar^2 
\mathcal{K}_x q_x}{m}
\nonumber \\  &
\pm\sqrt{\left(\zeta_{q_x}+2{\hbar\Omega}
\right)\left(\zeta_ { q_x}+2{\hbar\Omega}+2gn\right)} \,.
 \label{eq:dispersionM4_0}
\end{align}
These expressions are plotted in the bottom panel of Fig.~\ref{fig:M4disp} for 
a 
system with $\hbar\Omega=0.2 \, gn$ and $\mathcal{K}_x \,\xi=0.1$.

\item $k=1 \,$: The chemical potential is $\mu_1=gn+{\hbar^2 
\mathcal{K}_x^2}/{2m}$, 
and $\alpha_{1,p}=\{0,0,0,0\}$,  $\beta_{1,p}=\{0\,,1,\,-1,\,0\}$. The wave 
functions are:  
\begin{align}
\psi_{1,1}&=\,-\psi_{3,1}=i\sqrt{n}\, \exp{[i(\mathcal{K}_x\,x-\mu_1 
t/\hbar)]} \,,\nonumber\\
\psi_{4,1}&=\,-\psi_{2,1}=\sqrt{n}\, \exp{[i(\mathcal{K}_x\,x-\mu_1
t/\hbar)]} \,,
\label{eq:stable_vortex}
\end{align}
which yield a discrete, transverse vortex of circulation $\Gamma_1= 
4\hbar/m$. The interesting property of this state is its stability, 
irrespective of the axial momentum, which allows for its experimental 
realization. The dispersion curves $\omega_{1,p}$ contain
only real frequencies (see Fig.~\ref{fig:M4disp}):
\begin{align}
\hbar \omega_{1,0}=\hbar \omega_{1,2}=\frac{\hbar^2 \mathcal{K}_x q_x}{m}
\pm\sqrt{ \zeta_{q_x}\left(\zeta_{q_x}+2gn\right)} \,,\\
\hbar\omega_{1, 1}=\hbar\left(\frac{\hbar 
\mathcal{K}_x q_x}{m}+\Omega\right)\pm\sqrt{\zeta_{q_x}\left(\zeta_{q_x}+
2gn\right)} \,,\\
\hbar\omega_{1,-1}=\hbar\left(\frac{\hbar 
\mathcal{K}_x q_x}{m}-\Omega\right)\pm\sqrt{\zeta_{q_x}\left(\zeta_{q_x}+
2gn\right)} \,.
 \label{eq:dispersionM4_1}
\end{align}
We have also performed numerical simulations of the real time evolution of 
these states for $\hbar\Omega=0.2 \, gn$ and $\mathcal{K}_x \,\xi=0.1$ (same 
parameters as in Fig.~\ref{fig:M4disp}). Our numerical results obtained from 
the solution of the GP Eqs. (\ref{eq:Mtgp_pbc}), after seeding a random 
perturbation in the stationary state, confirm the dynamical stability of this 
state, since the initial configuration (\ref{eq:stable_vortex}) keeps robust 
against the perturbations. 

\item $k=2 \,$:  This state lies at the edge of 
the Brillouin zone, having maximum chemical potential $\mu_2=gn+{\hbar^2 
\mathcal{K}_x^2}/{2m}+\hbar\Omega$, and 
parameters $\alpha_{2,p}=\{0\,,-1,\,-1,\,-2\}$ and $\beta_{2,p}=\{0,0,0,0\}$. 
The wave function in each BEC component is:
\begin{align}
\psi_{1,2}&=&\psi_{3,2}=\sqrt{n}\,\exp{[i(\mathcal{K}_x\,x-\mu_2 t/\hbar)]} 
\,,\\ 
\psi_{2,2}&=&\psi_{4,2}= -\sqrt{n}\,\exp{[i(\mathcal{K}_x\,x-\mu_2 t/\hbar)]} 
\,.
\end{align}
In this state the Josephson circulation vanishes, $\Gamma_2= 0$, and the system 
presents a 
sequence of $\pi-$Josephson junctions which are unstable. This feature is 
captured by the linear dispersion, which shows several unstable branches
\begin{align}
\hbar \omega_{2,0}&=\frac{\hbar^2 
\mathcal{K}_x q_x}{m}\pm\sqrt{ \zeta_{q_x} 
\left(\zeta_{q_x}+2gn\right)} \,,\\
\hbar\omega_{2,\pm 1}&=\frac{\hbar^2 \mathcal{K}_x 
q_x}{m}
\\ \nonumber &
\pm\sqrt{\left(\zeta_{q_x}-\hbar\Omega\right)\left(\zeta_{q_x}
-\hbar\Omega + 2gn\right) } \,,\\
\hbar\omega_{2,2}&=\frac{\hbar^2 \mathcal{K}_x 
q_x}{m}
\nonumber \\  &
\pm\sqrt{\left(\zeta_{q_x}-2\hbar\Omega\right)\left(\zeta_{q_x}
-2\hbar\Omega + 2gn\right) } \,.
 \label{eq:dispersionM4_2}
\end{align}
Again our numerical simulations with the time-dependent GP 
Eqs.~(\ref{eq:Mtgp_pbc}), for such a 
state with $\hbar\Omega=0.2 \, gn$ and $\mathcal{K}_x \,\xi=0.1$, confirm the 
linear prediction and show the decay of the initial, constant density state.
\end{itemize}

\section{Nonlinear dynamics of localized states}
\label{sec:localized}

\begin{figure}[tb]
 \centering
  \includegraphics[width=0.9\linewidth]{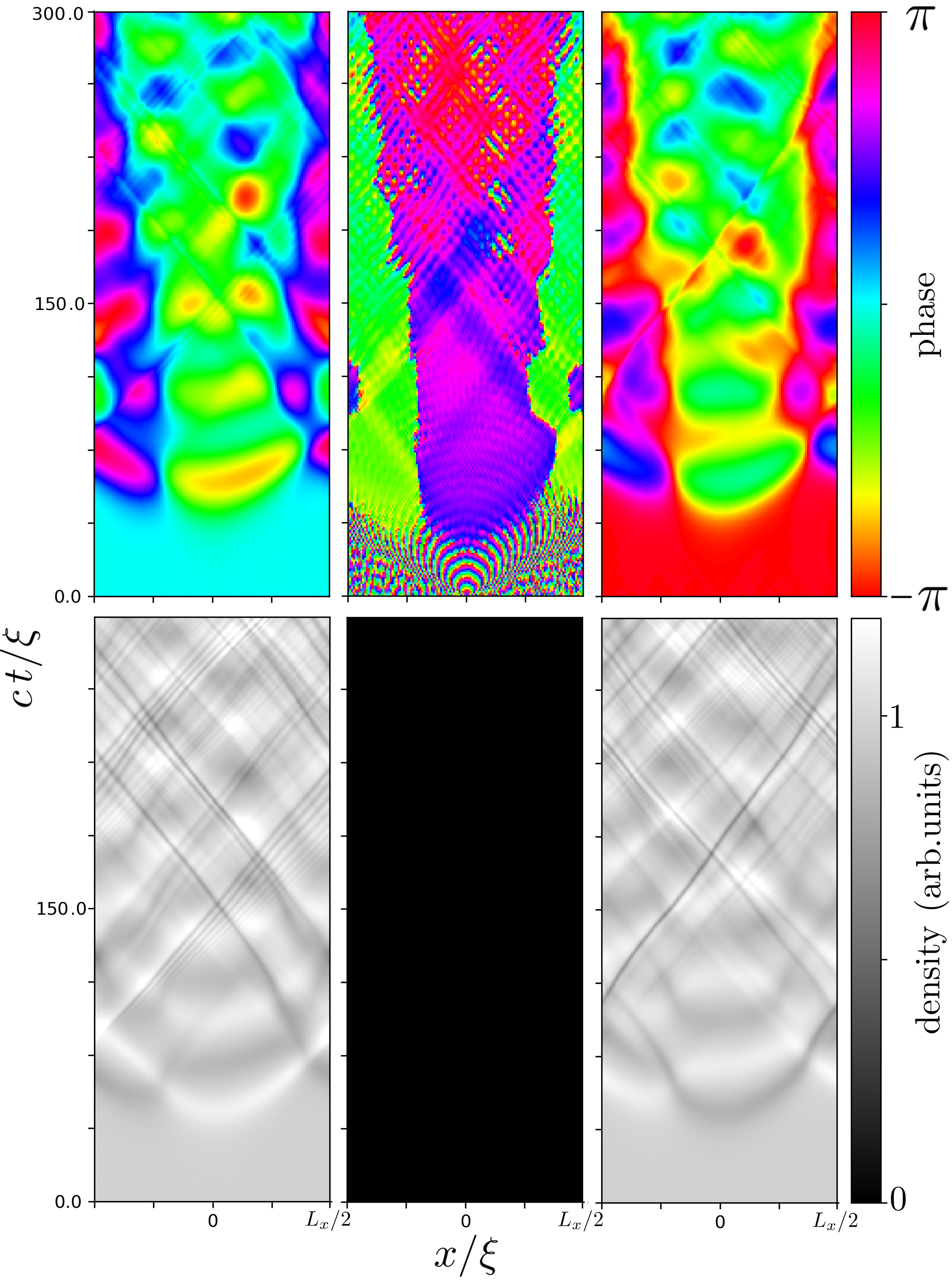}
 \caption{Time evolution of an antisymmetric state of the family $|k|=1$ and 
zero axial momentum in a $M=3$ stack of axial length 
$L_x=155\,\xi$ with periodic boundary conditions. Despite the small coupling, 
$\hbar \Omega=4\times10^{-3} \, gn$, the system is dynamically unstable. The 
BECs $j=$1, 2, 3 correspond to the panels from left to right. The initial 
state has been seeded with a random Gaussian perturbation (less than 1$\%$ in 
amplitude). }
 \label{fig:evolM3_k1a}
\end{figure}
\begin{figure}[tb]
 \centering
  \includegraphics[width=0.9\linewidth]{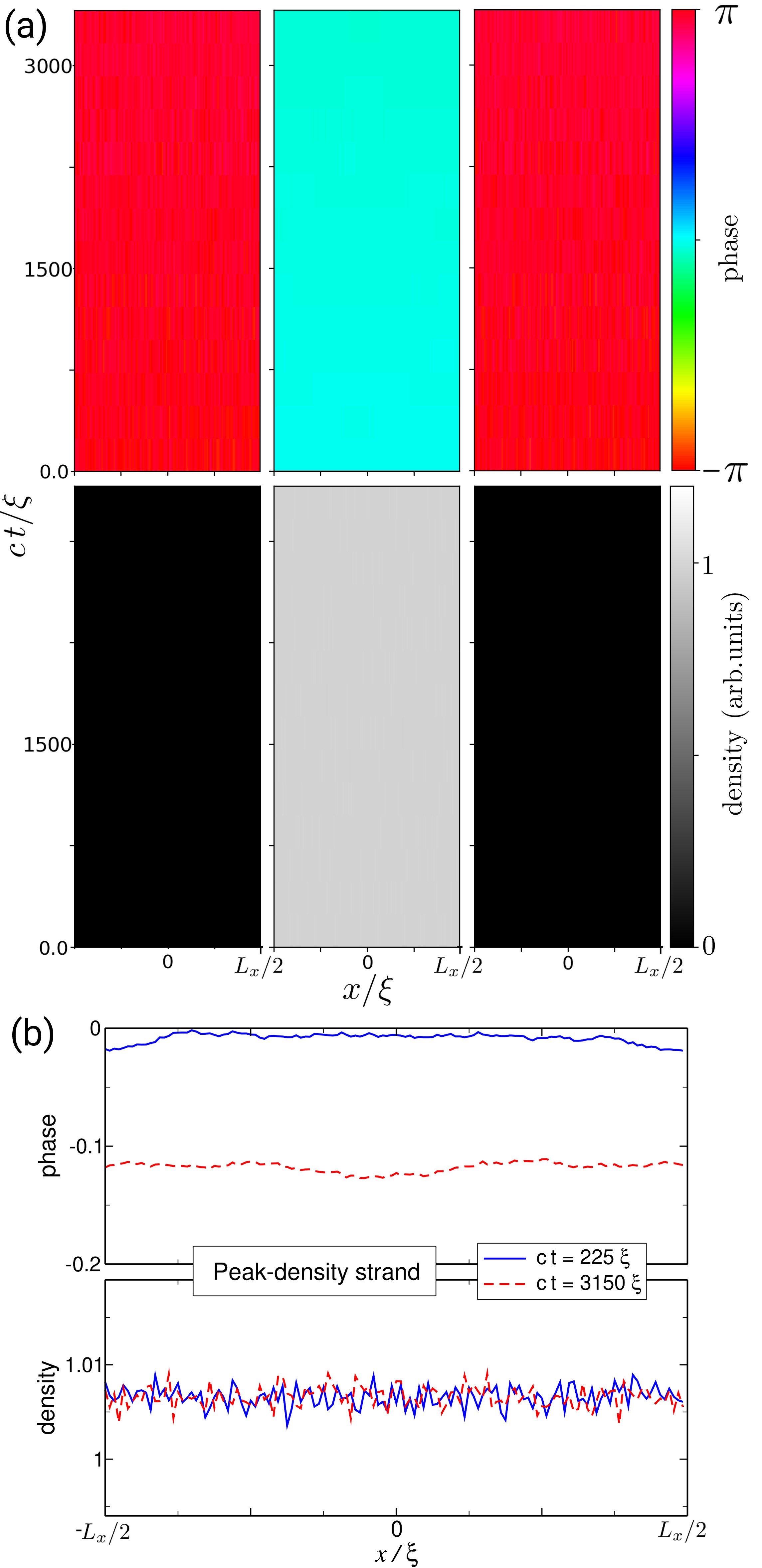}
 \caption{Time evolution of a stable, one-peak-symmetric state with 
zero axial momentum in a $M=10$ stack with the same parameters as the case of 
Fig.~\ref{fig:evolM3_k1a}. {\color{black} (a) The phase (top) and 
density (bottom) of the peak-density strand (middle) and the two 
adjacent strands (sides) are shown. (b)Two snapshots at 
different times of the local density (bottom panel) and the local phase (top 
panel) of the peak-density strand during the time evolution with initial 
perturbative noise.}
}
 \label{fig:evolM10_gap}
\end{figure}
We study the nonlinear dynamics of the localized states with one or two 
density peaks in the stack of constant density BECs. First, we
numerically solve the Bogoliubov equations in order to check the linear 
stability of the corresponding stationary state. Next, we perform the 
real time evolution with the GP Eq. 
(\ref{eq:Mtgp_pbc}) of this state after adding perturbative noise.

{\color{black} As predicted by the linear analysis of Sect. 
\ref{sec:localized_stab}, in the simplest stack with $M=3$, the nonlinear 
dynamics of the antisymmetric states 
is unstable.} To illustrate a typical decay process, Fig.~\ref{fig:evolM3_k1a} 
shows the real time evolution of an antisymmetric state with small coupling, 
$\hbar\Omega=4\times10^{-3} \, gn$, and zero axial momentum, $\mathcal{K}_x=0$. 
The data have been obtained from the numerical solution of GP 
Eq.~(\ref{eq:Mtgp_pbc}) 
with periodic boundary conditions in the axial coordinate. 
The axial length is $L_x=155\,\xi$.  
As can be seen, the initial nodal strand (middle panels in 
Fig.~\ref{fig:evolM3_k1a}) 
remains unpopulated during the whole 
evolution, and its phase is essentially undefined. The decay process is 
qualitatively different to the Bloch wave case presented in 
Fig.~\ref{fig:evolM3_k1}. 
The asymmetric state shows robust features of structural 
stability, roughly keeping {\color{black} the initial} density pattern across 
the stack.

On the contrary, {\color{black} we have checked that the nonlinear evolution 
of} 
the symmetric state with $M=3$, for the same parameters used 
above, is stable against perturbations. For larger stacks (we have performed
simulations up to $M=11$), our numerical results show that both the 
antisymmetric states (with two density 
peaks) in stacks with $ M\ge 4$, and the symmetric states (with 
one density peak) in stacks with $M\ge 3$ are also stable for the mentioned 
small coupling. However, the stability is lost at higher coupling values (at 
$\hbar \Omega\gtrsim 1\times10^{-2} \, gn$ for the 
parameters mentioned before). As a case example of stability, the time 
evolution of a symmetric state in a stack with $M=10$ components is shown in 
Fig.~\ref{fig:evolM10_gap}. {\color{black}In the top panels (a),} only the 
peak-density strand and its nearest neighbours are shown, since the other 
components have a practically null density. The initial, $t=0$, state has been 
seeded with perturbative noise, {\color{black} the detailed evolution of which 
at intermediate times is depicted on the bottom panels (b) for the peak-density 
strand.} As can be seen, the initial localized configuration is robust against 
the perturbations. Due to the strong density localization, the dynamics is 
insensitive to the change in the boundary conditions. Our results show that a 
one-peak state with open-boundary conditions follows a dynamics which is 
indistinguishable from that shown in  Fig.~\ref{fig:evolM10_gap}.

\vspace{.5cm}
\section{Discussion and conclusions}

The rich phenomenology presented by the stacks of parallel Josephson junctions 
can be readily realized in ultracold atomic gases by means of 1D or 2D optical 
lattices \cite{Budich2017,Baals2018}. These systems support nonlinear states 
whose dynamics reflects the interplay of continuous (along the axial 
$x$-direction of the BECs) and discrete (across the stack) features, and are 
promising candidates for pursuing technical applications with close
similarities to superconducting {\color{black} and photonic} devices.
In this work, we have contributed to this goal and have demonstrated the 
existence and stability of simultaneous superfluid currents flowing through 
both directions of a 2D stack. While the translation invariance along the 
$x$-axis allows for the excitation of axial-momentum eigenstates, the periodic 
arrangement of 
Josephson junctions induced by the linear coupling permits transverse Bloch 
waves carrying tunneling supercurrents. If the stack shapes a closed loop, 
these Josephson currents around it yield non-regular vortices whose 
circulation is a generic non-integer multiple of $h/m$. 

The dispersion relations of the transverse Josephson vortices have been 
obtained from the analytical solution of the linear Bogoliubov equations for 
the condensate excitations, and compared against the nonlinear time evolution 
of these states as given by the numerical solution of the Gross-Pitaeskii 
equation. In all the cases, the subsequent nonlinear dynamics is consistent 
with the stability predictions of the linear anaysis. 

For the sake of comparison with the usual coupled-sine-Gordon-equation model 
for coupled superconductors, a further linear analysis of the transverse 
Josephson vortices has been performed in the hydrodynamic limit. As a result,
we have derived linear wave-like equations for the relative phases and 
densities of the BEC components that resemble the 
mentioned model in the limit of small coupling.   

We have also shown that the Josephson supercurrents are suppressed in steady 
states that break the symmetry of the discrete lattice and can present a strong
localization across the stack. These nonlinear states belong to continues 
families of solutions to the Gross-Pitaevskii equation that can be tracked up 
to the non-interacting regime, where they are linear superposition of 
degenerate Bloch waves with opposite quasimomentum. Among these families, the 
gap-soliton-like states showing one or two dominant density peaks find
dynamical stability in finite systems within a small coupling regime.

The exploration of different topologies in the stack, or the effect of 
exposing the system to synthetic gauge fields \cite{Budich2017}, stand out 
as interesting ways of extending the present work that will be reported 
elsewhere.

\appendix*
\section*{Appendix: Long-wavelength excitations. Hydrodynamic 
approach}

We start by introducing low energy  perturbations [$\delta n_j(x,t), 
\delta\theta_j 
(x,t)$] around the density and  the phase of an equilibrium 
state  $\Psi_j=\sqrt{n_j}\exp(i\theta_j) \longrightarrow 
\sqrt{n_j+\delta n_j}\exp(i\theta_j+i\delta\theta_j)$. Then, we 
substitute the perturbed states in Eqs.~(\ref {eq:Mcontinuity}) and (\ref 
{eq:Mmomentum}), and keep terms up to first order in the perturbations.
We focus on the analysis of Bloch states with $n_j=n$. The mentioned procedure 
leads to
\begin{align}
&\frac{1}{n}\frac{\partial \delta n_j}{\partial t} = -\frac{\hbar}{m} 
\frac{\partial^2 \delta \theta_j}{\partial x^2}  \nonumber\\  &
-\Omega \left(\alpha_k (\sin\delta\theta_{j+1,j}-\sin\delta\theta_{j,j-1} )  + 
\beta_k \frac{\delta\bar n_{j+1,j}-\delta\bar n_{j,j-1}}{2n}\right),
\label{eq:dens_pert}\\
&\frac{\partial \delta\theta_j}{\partial t}= 
 - \frac{g \delta n_j}{\hbar} \label{eq:phase_pert} \\ \nonumber &
-\frac{\Omega}{2}\left(\beta_k(\sin\delta\theta_{j+1,j}
+\sin\delta\theta_{j,j-1}) -\alpha_k\frac {\delta n_{j+1,j} - \delta 
n_{j,j-1}}{2n}\right) \,, 
\end{align}
where $\delta \theta_{lj}=\delta \theta_l-\delta \theta_j$, $\delta 
n_{lj}=\delta n_l-\delta n_j$, $\delta \bar n_{lj}=\delta n_l+\delta 
n_j$ are the perturbations in relative phase, relative density and total
density, respectively, and $\alpha_k=\cos(2\pi\,k/M)$, 
$\beta_k=\sin(2\pi\,k/M)$. As usual in a long-wavelength
approximation, we have dropped the quantum-pressure term in 
Eq.~(\ref{eq:phase_pert}). 
For reasons that will become apparent later, we have 
kept the sine functions ($\sin\delta\theta_{lj}$) even in the linear 
approximation in order to track the Josephson currents, but they will be 
replaced by their argument (for consistency within the assumed 
first order approximation) at intermediate steps of the analytical derivations.

In what follows, we use the short notation $\rho_j=\delta n_{j+1,j}/n$,  
$\bar\rho_j=\delta \bar n_{j+1,j}/n$, $\phi_j=\delta \theta_{j+1,j}$, and also 
$\bar \phi_j=\delta\theta_{j+1}+\delta\theta_{j}$ for the total phase.
Since these quantities appear explicitly in previous expressions, we look 
for their equations of motion by adding and subtracting 
Eqs.~(\ref{eq:dens_pert}) 
and Eq.~(\ref{eq:phase_pert}) for consecutive components.
For the relative quantities we get
\begin{align}
\frac{\partial \rho_{j}}{\partial t} &= -\frac{\hbar}{m}  \!\! \left[ \!
\frac{\partial^2 \phi_j}{\partial x^2} + \!\alpha_k\frac{\delta^2 
\sin\phi_j}{\delta y^2} + \!\beta_k\frac{\rho_{j+1}-\rho_{j-1}}{2\,\delta 
y^2}\right] 
\label{eq:rdens_pert}
\\
\frac{\partial \phi_{j}}{\partial t} & = -\frac{m\,c^2}{\hbar}\rho_j + 
\nonumber  \\ &
\frac{\hbar}{m}\left[\frac{\alpha_k}{4}\frac{\delta^2 \rho_{j}}{\delta y^2}
-\beta_k \frac{\sin\phi_{j+1}-\sin\phi_{j-1}}{2\,\delta y^2}\right] \,, 
\label{eq:rphase_pert}
\end{align}
where the discrete operator $\delta^2 $ acts as 
$\delta^2 f_j=f_{j+1}-2f_j+f_{j-1}$. Exactly the same equations are obtained 
for the total quantities substituting $\rho$ by $\bar\rho$  and $\phi$ by 
$\bar\phi$.

As can be seen, relative and total quantities are decoupled in pairs of 
equations (\ref{eq:rdens_pert})-(\ref{eq:rphase_pert}). Within each pair,
by taking the time derivative of one of the equations and making use of the 
others, wave-like equations are obtained:
\begin{align}
\frac{1}{c^2}\frac{\partial^2 \rho_{j}}{\partial t^2} 
-\left((1+\hat\Omega\alpha_k)\frac{\partial^2}{\partial x^2} 
+(\alpha_k+\hat\Omega)\frac{\delta^2 }{\delta y^2}\right) \rho_{j}
\nonumber\\ +\frac{\hat\Omega}{2}\left(
\alpha_k \frac{\partial^2 }{\partial x^2}
+ (\alpha_k^2-\beta_k^2)\frac{\delta^2 }{\delta 
y^2}\right)(\rho_{j+1}+\rho_{j-1}) 
\nonumber\\ -2\hat\Omega\beta_k\left(\frac{\partial^2 }{\partial x^2}+\alpha_k
\frac{\delta^2 }{\delta y^2}\right)(\phi_{j+1}-\phi_{j-1}) =0 \,,
\label{eq:rdens_pert_long}
\\
\frac{1}{c^2}\frac{\partial^2 \phi_{j}}{\partial t^2} 
-\left((1+\hat\Omega\alpha_k)\frac{\partial^2}{\partial x^2} 
+(\alpha_k+\hat\Omega)\frac{\delta^2 }{\delta y^2}\right) \phi_{j}
\nonumber\\ +\frac{\hat\Omega}{2}\left(
\alpha_k \frac{\partial^2 }{\partial x^2}
+ (\alpha_k^2-\beta_k^2)\frac{\delta^2 }{\delta 
y^2}\right)(\phi_{j+1}+\phi_{j-1})
\nonumber\\ -\beta_k\left( 1-\frac{\hat\Omega}{2}\alpha_k\,\delta^2 \right)
\frac{\rho_{j+1}-\rho_{j-1}}{\delta y^2} =0 \,,
\label{eq:rphase_pert_long}
\end{align}
where $\hat\Omega=\hbar\Omega/2gn$. {\color{black} This system of M pairs of 
equations describes the linear dynamics of the BEC stack in the limit of 
long-wavelength excitations.}

\begin{acknowledgments}

M. G. and X. V.  acknowledge financial support from Ministerio de Econom{\'i}a 
y Competitividad (Spain), Agencia Estatal de Investigaci\'on (AEI) and Fondo 
Europeo de Desarrollo Regional (FEDER, EU) under Grants FIS2017-87801-P  and 
FIS2017-87534-P, from Generalitat de Catalunya Grant No. 2017SGR533,
and Project MDM-2014-0369 of ICCUB (Unidad de Excelencia María de Maeztu).
\end{acknowledgments}

\bibliography{three_components_v7iop_arxiv}

\end{document}